\documentclass[10pt]{IEEEtran}
\IEEEoverridecommandlockouts
\usepackage{bm}
\usepackage{algorithm, algorithmic}
\usepackage{setspace}
\usepackage{cite}
\usepackage{amsmath,amssymb,amsfonts}
\usepackage{algorithmic}
\usepackage{graphicx}
\usepackage{textcomp}
\usepackage{xcolor}
\usepackage{subfigure}
\usepackage{array}
\usepackage{CJK}
\newtheorem{lemma}{\hspace{0pt}\bf Lemma}
\def\BibTeX{{\rm B\kern-.05em{\sc i\kern-.025em b}\kern-.08em
    T\kern-.1667em\lower.7ex\hbox{E}\kern-.125emX}}
\begin{document}

\title{Joint User Identification, Channel Estimation, and Signal Detection for Grant-Free NOMA}

\author{Shuchao~Jiang, Xiaojun~Yuan, \textit{Senior Member, IEEE}, Xin~Wang, \textit{Senior Member, IEEE}, \\
        Chongbin~Xu, \textit{Member, IEEE}, and Wei~Yu, \textit{Fellow, IEEE}

\thanks{Manuscript received December 6, 2019; revised May 4, 2020; accepted June 20, 2020. The work of S. Jiang, X. Wang and C. Xu was supported in part by the National Natural Science Foundation of China (Grant No. 61501123 and 61671154). The work of X. Yuan was supported in part by The Key Areas of Research and Development Program of Guangdong Province, China, under Project 2018B010114001. The work of X. Wang and W. Yu was supported in part by the Innovation Program of Shanghai Municipal Science and Technology Commission Grant 17510710400. In addition, the work of W. Yu was supported in part by the Natural Sciences and Engineering Research Council of Canada. This paper was presented in part at IEEE GLOBECOM 2019, Hawaii, USA, December 2019 [40]. (Corresponding author: Chongbin Xu.) }
\thanks{Shuchao~Jiang, Xin~Wang, and Chongbin~Xu are with the Key Laboratory for Information Science of Electromagnetic Waves (MoE), Shanghai Institute for Advanced Communication and Data Science, Department of Communication Science and Engineering, Fudan University, Shanghai 200433, China (e-mail: \{17110720042, xwang11, chbinxu\}@fudan.edu.cn).}
\thanks{Xiaojun~Yuan is with the Center for Intelligent Networking and Communication (CINC),  University of Electronic Science and Technology of China, Chengdu 610000, China (e-mail: xjyuan@uestc.edu.cn).}
\thanks{Wei~Yu is with the Department of Electrical and Computer Engineering, University of Toronto, Toronto, Ontario M5S 3G4, Canada (e-mail: weiyu@comm.utoronto.ca).}}

\maketitle
\begin{abstract}
For massive machine-type communications, centralized control may incur a prohibitively high overhead. Grant-free non-orthogonal multiple access (NOMA) provides possible solutions, yet poses new challenges for efficient receiver design. In this paper, we develop a joint user identification, channel estimation, and signal detection (JUICESD) algorithm.
We divide the whole detection scheme into two modules: slot-wise multi-user detection (SMD) and combined signal and channel estimation (CSCE). SMD is designed to decouple the transmissions of different users by leveraging the approximate message passing (AMP) algorithms, and CSCE is designed to deal with the nonlinear coupling of activity state, channel coefficient and transmit signal of each user separately. To address the problem that the exact calculation of the messages exchanged within CSCE and between the two modules is complicated due to phase ambiguity issues, this paper proposes a rotationally invariant Gaussian mixture (RIGM) model, and develops an efficient JUICESD-RIGM algorithm. JUICESD-RIGM achieves a performance close to JUICESD with a much lower complexity. Capitalizing on the feature of RIGM, we further analyze the performance of JUICESD-RIGM with state evolution techniques. Numerical results demonstrate that the proposed algorithms achieve a significant performance improvement over the existing alternatives, and the derived state evolution method predicts the system performance accurately.
\end{abstract}
\begin{IEEEkeywords}
Grant-free NOMA, AMP, rotationally invariant Gaussian mixture (RIGM), state evolution
\end{IEEEkeywords}

\section{Introduction}
Massive machine-type communication (mMTC) is one of the most important scenarios for the next generation communications \cite{Hasan2013Random,Ghavimi2015M2M}. It is a key technology for realizing large-scale Internet of Things (IoT) applications such as smart home, smart manufacturing, and smart health care, etc. Different from conventional human-type communications, an mMTC scenario may involve a huge number of users. The packet arrival rate of each user can be low and the packet length is typically short \cite{Szabo2007Traffic}. In this case, the multiple access protocol plays a key role in supporting the massive connectivity efficiently \cite{Islam2014A,Xia2018Radio}. Due to the large signaling overhead, the conventional centralized control based multiple access techniques are generally not desirable.

Grant-free non-orthogonal multiple access (NOMA) has been proposed \cite{Nikopour2013Sparse,Dai2015Non,Liu2018Sparse} to reduce signaling overhead and enhance access capability. In grant-free NOMA, time and/or frequency domain resource blocks are divided into non-orthogonal sub-blocks that are shared by all potential users; and active users freely access the channel without waiting for any scheduling grant. This significantly reduces the overhead of control signaling to meet the requirement of mMTC.

Grant-free NOMA, however, poses a challenge for reliable receiver design. Besides channel estimation and signal detection, the receiver also needs to identify the activities of all potential users (i.e., which users simultaneously transmit packets) since there is no scheduling information at the receiver. A straightforward approach to the receiver design is first to identify active users, then to estimate the channel coefficients of the active users, and finally to recover the data of the active users. However, this separate processing approach may consume a substantial amount of spectrum and power resource, which in turn degrades the system performance.

For the sparse signal recovery problem involved in the receiver design of grant-free NOMA, compressed sensing (CS) \cite{Donoho2006Compressed} has been widely used. A possible approach for solving the CS problem is the $l_1$-norm relaxation via convex programming \cite{Tibshirani1996Regression}. However, the complexity of convex programming is high, especially for recovering high-dimensional signals. Other approximate algorithms have been proposed for more efficient sparse signal recovery, including match pursuit \cite{Mallat1993matching}, orthogonal match pursuit \cite{Tropp2004Greed}, iterative soft thresholding \cite{Figueiredo2007Gradient}, compressive sampling matching pursuit \cite{Needell2009CoSaMP}, approximate message passing (AMP) \cite{Donoho2009Message} and its variants \cite{Ma2014Turbo,Ma2016Orthogonal,RanganVector,Xue2017Denoising,Rangan2012Generalized,Rangan2016Fixed,Jason2014Bilinear,Parker2014Bilinear}. In particular, AMP provides a low-cost yet asymptotically optimal solution for a linear system with an independent and identically distributed (i.i.d.) sensing matrix \cite{Donoho2009Message}, and its performance can be accurately characterized by the state evolution \cite{Bayati2011The}. Furthermore, sparse signal recovery algorithms for more general system models have been developed recently, including turbo compressed sensing \cite{Ma2014Turbo}, orthogonal AMP (OAMP) \cite{Ma2016Orthogonal} and vector AMP (VAMP) \cite{RanganVector} for linear systems with a non-i.i.d. sensing matrix, generalized AMP (GAMP) \cite{Rangan2012Generalized,Rangan2016Fixed} for systems with nonlinear output, and bilinear GAMP (BiGAMP) \cite{Jason2014Bilinear,Parker2014Bilinear} for bilinear systems. These message passing based algorithms provide the current state of the art for sparse signal reconstruction.

Based on aforementioned algorithms, joint designs of channel estimation, user identification, and/or signal detection have been pursued to improve the system performance. Specifically, under the assumption of perfect channel state information (CSI) at the receiver (CSIR), joint active user identification and signal detection algorithms were developed in \cite{Wang2016Joint,Wei2017Approximate}. For systems without CSIR, \cite{Zhang2018Block,Liu2018Massive,Xu2015Active,Gabor2015Joint,Chen2018Sparse} established joint channel estimation and active user identification algorithms, followed by separated signal detection. In addition, joint channel and data estimation algorithms were developed for massive MIMO systems \cite{Wen2016Bayes,Zhang2017One,Zhang2018Blind,Ding2018Sparsity} and for single carrier systems \cite{Sun2018Joint}.

Recently, \cite{Du2018Joint} proposed a joint channel estimation and multiuser detection algorithm, named block sparsity adaptive subspace pursuit (BSASP). This algorithm transfers the single-measurement-vector compressive sensing (SMV-CS) problem to multiple-measurement-vector compressive sensing (MMV-CS), and reconstructs the sparse signal by exploiting the inherent block sparsity of the channel. BSASP generally suffers from the non-orthogonality of the training matrix. This issue becomes more serious in massive connectivity case, since the non-orthogonal training is usually inevitable in order to accommodate as many potential users as possible. In \cite{Fan2017Message}, a message-passing based joint channel estimation and data decoding algorithm was proposed for grant-free sparse code multiple access (SCMA) systems, where the messages were approximated by Gaussian distributions with minimized Kullback-Leibler (KL) divergence.

In this paper, we develop a joint user identification, channel estimation, and signal detection (JUICESD) algorithm based on message passing principles for grant-free NOMA systems. For this joint detection problem, the channel coefficients, the user activity states, and the transmit signals are coupled together, forming a complicated trilinear signal model to which the existing AMP algorithms (mostly developed for linear models) cannot be applied directly. Existing works mainly focus on some simplified signal models and only solve the problem partially, e.g., by assuming that the CSI is perfectly known at the receiver \cite{Wang2016Joint,Wei2017Approximate}, or by dividing the whole scheme into two phases, i.e., one phase for joint channel estimation and active user identification, and the other for separated signal detection \cite{Zhang2018Block,Liu2018Massive,Xu2015Active}. Different from the existing approaches, the main contributions of this paper include the following three aspects:
\begin{itemize}
\item By introducing appropriate auxiliary variables,we divide the whole detection scheme into two modules: slot-wise multi-user detection (SMD) and combined signal and channel estimation (CSCE). For SMD, which is designed to decouple the transmissions of different users, an AMP-type algorithm is developed to leverage the low complexity and the asymptotic optimality of AMP. For CSCE, which is designed to deal with the nonlinear coupling of the activity state, the channel coefficient and the transmit signal of each user, a message passing algorithm is derived in a user-by-user fashion. The overall algorithm, termed the JUICESD algorithm, is thus developed. It is shown that JUICESD achieves a significant performance improvement over the existing alternatives, and can even outperform linear minimum mean square error (LMMSE) receivers with oracle user activity information.

\item The exact calculation of the messages exchanged within the CSCE module and between the two modules in JUICESD involves computational complexity exponential in the frame length. To reduce computational complexity, we show that the messages in the JUICESD algorithm exhibit a rotational invariance property. We thus propose a rotationally invariant Gaussian mixture (RIGM) model for the message updates, and develop an efficient JUICESD-RIGM algorithm. JUICESD-RIGM achieves a performance close to JUICESD but with a much lower complexity that is quadratic in the frame length. Hence, it is well suited for machine type communications with massive devices and short packets.

\item Capitalizing on the feature of the proposed RIGM model, we further analyze the performance of JUICESD-RIGM by developing a state evolution technique. Numerical results show that the derived state evolution method predicts the system performance accurately. This analysis may provide useful insights for future system design and optimization.
\end{itemize}

The rest of this paper is organized as follows. Section \ref{sec.System} outlines the system model and formulates the problem of interest. Section \ref{sec.JUICESD} develops the JUICESD algorithm, while Section \ref{sec.JUICESD-CAGMA} proposes the JUICESD-RIGM algorithm. The performance of JUICESD-RIGM is characterized by using state evolution techniques in Section \ref{StateEvolution}. Numerical results are provided in Section \ref{sec.Numerical}. Section \ref{sec.Conclusion} concludes the paper.

{\em{Notation}}: We use boldface uppercase letters such as $\bm{A}$ to denote matrices, and use boldface lowercase letters such as $\bm{b}$ to denote vectors; $\bm{a}_k$ denotes the $k$-th column of matrix $\bm{A}$, $a_{l,k}$ denotes the entry in the $l$-th row and $k$-th column of matrix $\bm{A}$; and $b_k$ denotes the $k$-th element of vector $\bm{b}$. For matrices and vectors, $(\cdot)^T$ denotes transpose, ${\rm diag}(\bm{b})$ is the diagonal matrix with the diagonal elements specified by $\bm{b}$. Denote by $\cal A$ a set, and by $\vert \cal A \vert$ the cardinality of $\cal A$; $\mathcal{N}(\mu,\tau)$ denotes the Gaussian distribution with mean $\mu$ and variance $\tau$; $\mathcal{CN} (\mu,\tau)$ denotes the complex Gaussian distribution with mean $\mu$, variance $\tau$, and zero relation.
\section{System Model}\label{sec.System}
Consider a typical mMTC scenario, in which a large number of single-antenna users with sporadic traffic communicate with a single-antenna access point (AP) \cite{Wei2017Approximate,Du2018Joint,Chen2018Sparse,Gabor2015Joint,Ahn2019EP}. Based on the received signals, the AP is responsible for judging which users are active, estimating the channels of the active users, and recovering the signals transmitted by the active users.
\subsection{Grant-Free NOMA Transmission}
We follow the spreading based NOMA schemes in \cite{Wei2017Approximate,Wang2016Joint,Du2018Joint} for grant-free transmissions. Specifically, each user $k$ is assigned with a unique spreading sequence $\bm{a}_k = [a_{1,k}, \cdots, a_{L,k}]^T$ as its signature, where $L$ is the spreading length. For massive connectivity case, $L$ can be much less than the total number of users $K$, and therefore orthogonal spreading sequence design is generally impossible. Here we assume that the elements of each $\bm{a}_k$ are randomly and independently drawn from the Gaussian distribution $\mathcal N(0, 1/L)$.

Consider the transmission in a frame of $T$ slots, where each slot consists of $L$ transmission symbols corresponding to the length of the spreading sequence. The received signal can be modeled as \cite{Du2018Joint,Jiang2019}
\begin{equation}\label{eq.sys2}
\bm{r}_{t} = \sum_{k=1}^K \bm{a}_k h_k u_k x_{k,t} + \bm{w}_t, \ t = 1, \cdots, T
\end{equation}
where $K$ is the total number of users, $h_k$ is the channel coefficient from user $k$ to the AP, $u_k$ is an indicator to represent the activity state of user $k$ (with $u_k=1$ meaning that user $k$ is active and $u_k=0$ otherwise), $x_{k,t}$ is the transmit signal of user $k$ at time slot $t$, and each entry of $\bm{w}_t$ is the complex additive white Gaussian noise (AWGN) with mean zero and variance $N_0$. All $\{ h_k \}$, $\{u_k\}$, and $\{x_{k,t}\}$ are assumed to be independently distributed.

Block fading is assumed, i.e., $h_k$ and $u_k$ remain unchanged within each transmission frame. The channel coefficient $h_k$ is modelled as $h_k = \sqrt{\beta_k}{\alpha_k},\ \forall k$, where $\alpha_k \sim \mathcal{CN}(0,1)$ denotes the Rayleigh fading component, and $\beta_k$ denotes large scale fading component including path-loss and shadowing. Then we have $h_k \sim p_{H_k}( h_k )= \mathcal{CN}(0,\beta_k)$, implying that the channels of different users are not necessarily identically distributed.

\begin{figure*}
    \centering
    \includegraphics[width = 0.9\textwidth]{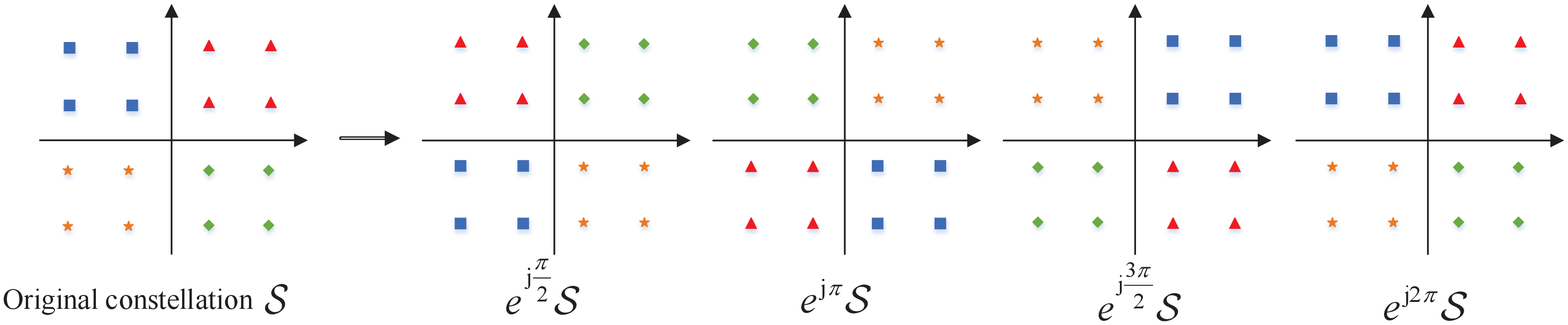}
    \caption{Rotational invariance of the standard 16QAM constellation.}\label{Fig.16QAM}
\end{figure*}

The user symbols $\{x_{k,t}\}$ are modulated by using a common signal constellation ${\cal S}$ with cardinality $\vert {\cal S} \vert$, i.e., each $x_{k,t}$ is randomly and uniformly drawn from ${\cal S}$. We say that ${\cal S}$ is rotationally invariant with respect to a phase shift $\theta$ if ${\cal S} = e^{ \text{j} \theta} {\cal S}$. The rotationally invariant set is defined as $\Omega_{\cal S}=\{ \theta| {\cal S} = e^{ \text{j} \theta} {\cal S}, \ 0<\theta \leq 2\pi \}$. By noting that $\Omega_{\cal S}$ is a cyclic group under addition, we can generally express $\Omega_{\cal S}$ as $\Omega_{\cal S} = \{ \theta_0, 2\theta_0, \cdots, |\Omega_{\cal S}|\theta_0 \}$, where $\theta_0$ is the minimum value in $\Omega_{\cal S}$. Such a rotational invariance property holds for commonly used modulation schemes such as phase shift keying (PSK) and quadrature amplitude modulation (QAM), and will be utilized later in the algorithm design in Section \ref{sec.JUICESD-CAGMA}. Fig. \ref{Fig.16QAM} shows the example of the standard 16QAM where ${\cal S} = \frac{1}{\sqrt{10}}\{x_r + \text{j} x_i \,|\, x_r, x_i \in \{-3, -1, 1, 3\} \}$, and $\Omega_{\cal S} = \{0.5 \pi, \pi, 1.5 \pi, 2 \pi \}$.

Throughout the paper, we make the following assumptions \cite{Liu2018Sparse,Du2018Joint}:
\begin{itemize}
  \item All users are synchronized in frames. This can be achieved by the AP sending a beacon signal to initialize uplink transmissions.
  \item The packet arrival rate of each user per frame is Bernoulli distributed with parameter $\lambda \in \left( 0,1 \right]$. Each user starts to transmit a packet at the beginning of a frame with probability $\lambda$.
  \item Retransmissions are not considered and all packets are regarded as new arrivals.
\end{itemize}

\subsection{Problem Formulation}
The focus of this paper is the design of an efficient receiver to identify the activity of all potential users, to estimate the channel coefficients, and to recover the transmit data of the active users. To this end, we rewrite the received signal (\ref{eq.sys2}) as
\begin{equation}\label{eq.SignalModelMatrix}
\bm{R} = \bm{AHUX} + \bm{W}
\end{equation}
%\begin{spacing}{1.3}
\noindent where $\bm{R} = [\bm{r}_1, \cdots, \bm{r}_T]\in\mathbb{C}^{L \times T}$, $\bm{A} = [\bm{a}_1, \cdots, \bm{a}_K]\in\mathbb{R}^{L \times K}$, $\bm{H} = {\rm diag}([h_1, \cdots, h_K]^T)$ $\in\mathbb{C}^{K \times K}$, $\bm{U} = {\rm diag}([u_1, \cdots, u_K ]^T)\in\{0,1\}^{K \times K}$, $\bm{X} = [x_{k,t}] \in {\cal S}^{K \times T}$, and $\bm{W} = [\bm{w}_1, \cdots, \bm{w}_T]\in\mathbb{C}^{L \times T}$.

Suppose that the AP jointly estimates ($\bm{H}$, $\bm{U}$, $\bm{X}$) by following the maximum \emph{a posteriori} probability (MAP) principle. Conditioned on $\bm{R}$, the \emph{a posteriori} probability density function (PDF) is given by
\begin{multline}\label{eq.prob1_pre}
p_{H, U, X|R}(\bm{\bm{H}, \bm{U}, \bm{X}|R}) \propto \\
{\rm exp} \left( -\frac{ \Vert \bm{R-AHUX} \Vert _F^2 }{N_0} \right)p_{H}(\bm{H})p_U(\bm{U})p_X(\bm{X})
\end{multline}

\noindent where the channel coefficient $h_k$ is drawn from $\mathcal{CN}(0,\beta_k)$; the activity indicator $u_k$ is drawn from the Bernoulli distribution $p_{u}(u_k) = (1-\lambda) \delta(u_k) + \lambda \delta(u_k-1)$; the elements of $\bm{X}$ are independently drawn from a uniform distribution over the signal constellation ${\cal S}$, i.e., $p_{\cal S}( s ) = \frac{1}{\vert {\cal S} \vert} \sum \nolimits_{j=1}^{\vert {\cal S} \vert} \delta(s-s_j)$ with $\sum \nolimits_{j=1}^{\vert {\cal S} \vert} \Vert s_j \Vert^2/\vert {\cal S} \vert = 1$ for power normalization.

Based on (\ref{eq.prob1_pre}), the MAP estimate of ($\bm{H}$, $\bm{U}$, $\bm{X}$) is given by
\begin{multline}\label{eq.q1}
\left( \hat{\bm{H}}, \hat{\bm{U}}, \hat{\bm{X}} \right) = \mathop{ \arg\max}_{(\bm{H,U,X})}   {\rm exp} \left( -\frac{ \Vert \bm{R-AHUX} \Vert _F^2 }{N_0} \right)  \\  \cdot p_{H}(\bm{H})p_U(\bm{U})p_X(\bm{X}) .
\end{multline}
The solution to problem (\ref{eq.q1}) is not unique. For example, let  $\bm{D} = {\rm diag} ( [ e^{- \text{j} \theta_1}, \cdots, e^{- \text{j} \theta_K} ]^{T} )$ with $\theta_{k} \in \Omega_{\cal S}$, $k=1,2,\cdots,K$. Then, from the rotational invariance property of the constellation ${\cal S}$, we see that $\bm{DX}$ has the same distribution as $\bm{X}$ does. Besides, $\bm{HD^{-1}}$ has the same distribution as $\bm{H}$ does, which is true for almost all the existing wireless random channel models. Thus, if $\left( \hat{\bm{H}}, \hat{\bm{U}}, \hat{\bm{X}} \right)$ is a solution to problem (\ref{eq.q1}), then $\left( \hat{\bm{H}}\bm{D^{-1}}, \hat{\bm{U}}, \bm{D} \hat{\bm{X}} \right)$ is also a valid solution to problem (\ref{eq.q1}). This phenomenon is referred to as the phase ambiguity of problem (\ref{eq.q1}), which appears widely when data and channels need to be estimated jointly. To remove the phase ambiguity, a simple approach is to insert reference symbols at the first (or any other) column of $\bm{X}$, i.e., each user needs at least one reference symbol for elimination of phase ambiguity.

Problem (\ref{eq.q1}) is non-convex and generally difficult to solve. Message passing algorithms could provide possible solutions. However, the variables to be estimated in (\ref{eq.q1}), i.e., $\bm{H}$, $\bm{U}$, and $\bm{X}$, are all coupled to form a trilinear function. Exact message passing based on the sum-product rule is too complicated to implement, while the existing low complexity AMP-type algorithms \cite{Donoho2009Message,Ma2014Turbo,Ma2016Orthogonal,RanganVector,Xue2017Denoising,Rangan2012Generalized,Rangan2016Fixed,Jason2014Bilinear,Parker2014Bilinear} cannot be applied to the trilinear model in (\ref{eq.SignalModelMatrix}) directly. To bypass the dilemma, we next develop a low-complexity yet efficient iterative algorithm to solve the problem by a judicious design of the receiver structure and the message updates.
\section{Proposed JUICESD Algorithm}\label{sec.JUICESD}
To facilitate a low-complexity yet efficient solution to the joint detection problem in (\ref{eq.q1}), we divide the whole detection scheme into two modules by introducing appropriate auxiliary variables, based on which we develop the proposed JUICESD algorithm.
\subsection{JUICESD Algorithm Structure}\label{AA}
Introduce the following auxiliary variables:
\begin{equation}\label{eq.gk}
g_k \doteq h_k u_k, \ \forall k,
\end{equation}
\begin{equation}\label{eq.ykt}
y_{k,t} \doteq g_k x_{k,t}, \ \forall k, t. %= h_k I_k x_{k,t}
\end{equation}
We henceforth refer to $g_k$ and $y_{k,t}$ as the effective channel of user $k$ and the effective signal of user $k$ at time slot $t$, respectively. Based on the \emph{a priori} distributions of $h_k$, $u_k$, and $x_{k,t}$, we obtain the \emph{a priori} distributions of $g_k$ and $y_{k,t}$ from (\ref{eq.gk}) and (\ref{eq.ykt}) respectively as
\begin{equation}\label{eq.pg}
%p_{G_k}(g_k)= (1-\lambda)\delta(g_k)+ \frac{\lambda}{\pi \tau_h} {\rm exp} \left( -\frac{\Vert g_k \Vert^2}{\tau_h} \right),
p_{G_k}(g_k)= (1-\lambda)\delta(g_k)+ \lambda p_{H_k}(g_k),
\end{equation}
\begin{equation}\label{eq.pykt}
%{p}_{y_{k,t}}(y_{k,t}) = (1-\lambda) \delta(y_{k,t}) + \frac{\lambda}{\vert {\cal S} \vert} \sum_{j=1}^{\vert {\cal S} \vert} \frac{1}{\pi \Vert s_j \Vert^2 \tau_h }{\rm exp} \left( -\frac{ \Vert y_{k,t} \Vert^2 }{\Vert s_j \Vert^2 \tau_h } \right).
{p}_{y_{k,t}}(y_{k,t}) = (1-\lambda) \delta(y_{k,t}) + \frac{\lambda}{\vert {\cal S} \vert} \sum_{j=1}^{\vert {\cal S} \vert} p_{H_k}(y_{k,t}/s_j).
\end{equation}

Rewrite the system model in (\ref{eq.sys2}) as
\begin{equation}\label{eq.syse}
\bm{r}_t = \sum_{k=1}^K \bm{a}_k y_{k,t} + \bm{w}_{t}, \ t = 1, \cdots, T.
\end{equation}
Clearly, the signal model (\ref{eq.syse}) is linear in the auxiliary variables $\{y_{k,t}\}$ since $\{\bm{a}_k\}$ are known to the receiver. In addition, given $\{y_{k,t}\}$, the estimations of $\{g_k\}$ and $\{x_{k,t}\}$ are nonlinear yet decoupled for different $k$. These two properties are of importance to our algorithm design.

\begin{figure}
    \centering
    \includegraphics[width = 0.5\textwidth]{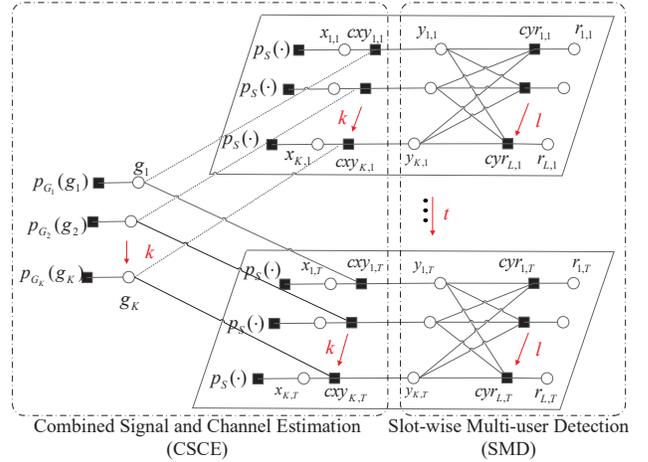}
    \caption{Factor graph representation of the considered system.}\label{Fig.FactorGraph}
\end{figure}

With (\ref{eq.ykt})-(\ref{eq.syse}), we can represent the system model by the factor graph in Fig. \ref{Fig.FactorGraph}. The factor graph consists of two types of nodes:
\begin{itemize}
  \item Variable nodes $\{g_k\}$,$\{x_{k,t}\}$, $\{y_{k,t}\}$, and $\{r_{l,t}\}$, depicted as white circles in Fig. \ref{Fig.FactorGraph}, corresponding to the variables in (\ref{eq.ykt}) and (\ref{eq.syse}), where $r_{l,t}$ is the $l$-th entry of $\bm{r}_t$;
  \item Check nodes $\{p_{G_k}(g_k)\}$, $p_{\cal S}( \cdot )$, $\{cxy_{k,t}\}$, and $\{cyr_{l,t}\}$, depicted as black boxes in Fig. \ref{Fig.FactorGraph}, corresponding to the marginal \emph{a priori} distributions of $\{g_k\}$, the marginal \emph{a priori} distributions of $\{x_{k,t}\}$, the equality constraints in (\ref{eq.ykt}), and the equality constraints in (\ref{eq.syse}), respectively.
\end{itemize}
A variable node is connected to a check node when the variable is involved in the check constraint.

As shown in Fig. \ref{Fig.FactorGraph}, we divide the whole receiver structure into two modules, one for the multi-user signal model in (\ref{eq.syse}) and the other for the nonlinear equality constraint in (\ref{eq.ykt}). We next outline their main functionalities.

The module on the right hand side of Fig. \ref{Fig.FactorGraph} focuses on the multi-user signal model in (\ref{eq.syse}). With the messages of $\{y_{k,t}\}$ fed back from the other module as the \emph{a priori}, the estimation of $\{y_{k,t}\}$ given $\bm{r}_t$ in (\ref{eq.syse}) can be performed slot-by-slot. Hence, we refer to this module as slot-wise multi-user detection (SMD). The refined estimates of $\{y_{k,t}\}$ are forwarded to the other module. The detailed operations will be described later in Section III-B.

The module on the left hand side of Fig. \ref{Fig.FactorGraph} is for the nonlinear constraints in (\ref{eq.ykt}). With the estimates of $\{y_{k,t}\}$ as input, the effective channels $\{g_k\}$ and the effective signals $\{y_{k,t}\}$ can be refined based on (\ref{eq.ykt}) by noticing that the effective channel of the same user remains unchanged in one frame. Thus we refer to this module as combined signal and channel estimation (CSCE). The output of the CSCE is forwarded to the SMD module for further processing. The detailed operations of the CSCE will be specified in Section III-C.

The above two modules are executed iteratively. Upon convergence, we obtain the final detection results as follows.

\emph{User identification}: Let $\hat{g}_k$ be the estimate of $g_k$ after the last iteration. Then, a user is active if the magnitude of $\hat{g}_k$ is larger than a certain predetermined threshold $g_{th}$, i.e.
\begin{equation}\label{eq.ua}
\hat{u}_k =
  \begin{cases}
     1  &| \hat{g}_k | \geq g_{th},\\
     0  &| \hat{g}_k | < g_{th}.
  \end{cases}
\end{equation}

\emph{Channel estimation}: Let $\hat{y}_{k,t}$ be the final estimate of $y_{k,t}$, $\forall k,t$. From (\ref{eq.ykt}), with $\{\hat{y}_{k,t}\}$ and $\hat{u}_k$ available, the channel coefficient of active user $k$ can be calculated from the reference symbol transmission (i.e., $\hat{y}_{k,1}$) as
\begin{equation}\label{eq.ce}
%\hat{h}_k = \frac{1}{T_p}\sum_{t=1}^{T_p} \frac{\hat{y}_{k,t}}{x_{k,t}}
\hat{h}_k = \hat{y}_{k,1} / {s_p}, \ \text{for} \ \hat{u}_k=1,
\end{equation}
where $s_p$ is the common reference symbol at the first column of $\bm{X}$.

\emph{Signal detection}: With the estimates $\hat{y}_{k,t}$ and $\hat{h}_k$ available, a soft estimate of the transmit signal $x_{k,t}$ (with $\hat{u}_k = 1$ and $2 \le t \le T$) is given by
\begin{equation}\label{eq.sd}
\hat {x}_{k,t} = \hat{y}_{k,t} / \hat{h}_k.
\end{equation}
Then a hard decision on $\hat{x}_{k,t}$ can be made accordingly.
\subsection{SMD Operation}
The SMD module is to estimate $\{y_{k,t}\}$ based on the received signal model (\ref{eq.syse}) and the messages of $\{y_{k,t}\}$ from the CSCE module. AMP-type algorithms can be used to provide a near-optimal estimation of $\{y_{k,t}\}$. Specifically, we follow the GAMP algorithm \cite{Rangan2012Generalized}. The messages passed between the nodes are approximated by Gaussian distributions, so that only the means and variances of the messages are involved in message exchanges. In addition, instead of calculating the messages on each edge, the GAMP algorithm calculates the messages on each node \cite{Bayati2011The,Rangan2012Generalized}; hence, the number of messages calculated in the algorithm can be significantly reduced.

We next outline the SMD operation in each time slot $t$ by following the GAMP algorithms \cite{Rangan2012Generalized,Rangan2016Fixed}. The detailed derivations are omitted for brevity.

\emph{1) Initialization:}
Denote by $\{ m_{cxy_{k,t} \rightarrow y_{k,t}}(y_{k,t}) \}$ the messages of $\{y_{k,t}\}$ fed back from CSCE. (Specifically, they are passed from the check nodes $\{cxy_{k,t}\}$ as detailed in Section III-C.) With no feedback from CSCE at the beginning, each $m_{cxy_{k,t} \rightarrow y_{k,t}}(y_{k,t})$ is initialized to ${p}_{y_{k,t}}(y_{k,t})$ in (\ref{eq.pykt}). The means $\{\hat{y}_{k,t}\}$ and variances $\{v_{y_{k,t}}\}$ of $\{y_{k,t}\}$ are calculated at variable nodes $\{y_{k,t}\}$.

\emph{2) Message update at check nodes $\{cyr_{l,t}\}$:} Based on the linear model $z_{l,t} = \sum\nolimits_{k=1}^{K} a_{l,k} y_{k,t}$, the messages of $\{y_{k,t}\}$ are cumulated to obtain an estimate of $\{z_{l,t}\}$. With the ``Onsager" correction applied, the messages of $\{z_{l,t}\}$ in the form of means $\{\hat{p}_{l,t}\}$ and variances $\{v_{p_{l,t}}\}$ are calculated as \cite{Rangan2012Generalized}
\begin{equation}\label{SMD_P_Variance}
v_{p_{l,t}} = \sum_{k=1}^{K} \vert a_{l,k} \vert^2 v_{y_{k,t}}, \ \forall l,
\end{equation}
\begin{equation}\label{SMD_P_Mean}
\hat{p}_{l,t} = \sum_{k=1}^K a_{l,k} \hat{y}_{k,t} - v_{p_{l,t}} \hat{s}_{l,t}, \ \forall l,
\end{equation}
where initially we set $\hat{s}_{l,t} = 0$ for $\forall l$. Then, the means $\{\hat{z}_{l,t}\}$ and the variances $\{v_{z_{l,t}}\}$ are computed by using the observations $\{r_{l,t}\}$ as
\begin{equation}\label{SMD_Z_Variance}
v_{z_{l,t}} = {\rm Var} \{ z_{l,t} | \hat{p}_{l,t}, v_{p_{l,t}},r_{l,t} \}, \ \forall l,
\end{equation}
\begin{equation}\label{SMD_Z_Mean}
\hat{z}_{l,t} = {\rm E}\{ z_{l,t} | \hat{p}_{l,t}, v_{p_{l,t}},r_{l,t} \}, \ \forall l,
\end{equation}
where the mean ${\rm E} \{ \cdot \}$ and variance ${\rm Var} \{ \cdot \}$ operations are taken with respect to the \emph{a posteriori} distribution of $z_{l,t}$ given the \emph{a priori} distribution $z_{l,t} \sim \mathcal{CN} (\hat{p}_{l,t},v_{p_{l,t}})$ and the observation $r_{l,t} = z_{l,t} + w_{l,t}$ with $w_{l,t}$ being the $l$-th entry of $\bm{w}_t$.
Lastly, the residual $\{\hat{s}_{l,t}\}$ and the inverse-residual-variances $\{v_{s_{l,t}}\}$ are computed by
\begin{equation}\label{SMD_S_Variance}
v_{s_{l,t}} = \left( 1- v_{z_{l,t}}/v_{p_{l,t}} \right) / v_{p_{l,t}} \ \forall l,
\end{equation}
\begin{equation}\label{SMD_S_Mean}
\hat{s}_{l,t} = \left(\hat{z}_{l,t} - \hat{p}_{l,t} \right) /v_{p_{l,t}} \ \forall l.
\end{equation}

\emph{3) Message update at variable nodes $\{y_{k,t}\}$:} With the residual $\{\hat{s}_{l,t} \}$ and inverse-residual-variances $\{v_{s_{l,t}}\}$, the messages of $\{y_{k,t}\}$ are computed in the form of means $\{\hat{r}_{k,t}\}$ and variances $\{v_{r_{k,t}} \}$ as
\begin{equation}\label{SMD_R_Variance}
v_{r_{k,t}}  = \left( \sum _{l=1}^{L} \vert a_{l,k} \vert^2 v_{s_{l,t}}  \right) ^{-1}, \ \forall k,
\end{equation}
\begin{equation}\label{SMD_R_Mean}
\hat{r}_{k,t}  = \hat{y}_{k,t}  + v_{r_{k,t}}  \sum_{l=1}^{L} a_{l,k} \hat{s}_{l,t} , \ \forall k.
\end{equation}
Then the means and variances of $\{y_{k,t}\}$ are updated by
\begin{equation}\label{SMD_Y_Variance}
v_{y_{k,t}} = {\rm Var} \{ y_{k,t} | m_{cxy_{k,t} \rightarrow y_{k,t}}(y_{k,t}), \hat{r}_{k,t} , v_{r_{k,t}}  \}, \ \forall k,
\end{equation}
\begin{equation}\label{SMD_Y_Mean}
\hat{y}_{k,t} = {\rm E}\{ y_{k,t} | m_{cxy_{k,t} \rightarrow y_{k,t}}(y_{k,t}), \hat{r}_{k,t} , v_{r_{k,t}}  \}, \ \forall k,
\end{equation}
where the mean ${\rm E} \{ \cdot \}$ and variance ${\rm Var} \{ \cdot \}$ operations are taken with respect to the \emph{a posteriori} distribution of $ y_{k,t} $ given its \emph{a priori} distribution $m_{cxy_{k,t} \rightarrow y_{k,t}}(y_{k,t})$ and the feedback message $y_{k,t} \sim \mathcal{CN} (\hat{r}_{k,t} ,v_{r_{k,t}} )$. The operations in (\ref{SMD_Y_Variance}) and (\ref{SMD_Y_Mean}) essentially give a nonlinear denoiser since they take the structure information of $y_{k,t}$ such as sparsity into account by regarding $m_{cxy_{k,t} \rightarrow y_{k,t}}(y_{k,t})$ as the \emph{a priori} distribution. The above refined messages are used to update the messages of $\{z_{l,t}\}$ in the next iteration. The iteration continues until convergence or the maximum iteration number $Q$ is reached. Finally, the messages of $\{y_{k,t}\}$ with mean $\{\hat{r}_{k,t} \}$ and variance $\{v_{r_{k,t}} \}$ are passed to the CSCE module.

\subsection{CSCE Operation}
With the messages $\{\hat{r}_{k,t} \}$ and $\{v_{r_{k,t}} \}$ from the SMD as input, the CSCE module in Fig. \ref{Fig.FactorGraph} deals with the nonlinear constraints in (\ref{eq.ykt}) to yield more accurate estimates of $\{y_{k,t}\}$. Since the constraints in (\ref{eq.ykt}) are decoupled for different users, we next present the message passing operations for each individual user as follows.

\emph{1) Messages passed from $\{cxy_{k,t}\}$ to $\{g_k\}$:} With the output of SMD, the message from $y_{k,t}$ to $cxy_{k,t}$ is given by a complex Gaussian distribution with mean $\hat{r}_{k,t} $ and variance $v_{r_{k,t}} $, i.e., $m_{y_{k,t} \rightarrow cxy_{k,t}}(y_{k,t}) = \mathcal{CN}(\hat{r}_{k,t} ,v_{r_{k,t}} )$. Then, from the sum-product rule, we obtain
\begin{equation}\label{eq.y2ht}
\begin{split}
&m_{cxy_{k,t} \rightarrow g_k}(g_{k}) \\
& = \int_{y_{k,t}, x_{k,t}} \hspace{-0.8cm}m_{y_{k,t} \rightarrow cxy_{k,t}}(y_{k,t}) p_{\cal S}(x_{k,t}) \delta(y_{k,t}-g_k x_{k,t}) \\
&= \sum_{j=1}^{\vert {\cal S} \vert} \frac{p_{\cal S}(s_j)}{\pi v_{r_{k,t}}  / \Vert s_j \Vert^2 } {\rm exp}\left(-\frac{\Vert g_{k} - \hat{r}_{k,t} /s_j \Vert^2}{v_{r_{k,t}}  / \Vert s_j \Vert^2}\right), \ \forall t,
\end{split}
\end{equation}
where $p_{\cal S}( x_{k,t} ) = \frac{1}{\vert {\cal S} \vert} \sum \nolimits_{j=1}^{\vert {\cal S} \vert} \delta(x_{k,t}-s_j)$ is the \emph{a priori} distribution of the transmit signal $x_{k,t}$.

In (\ref{eq.y2ht}), the algorithm assumes that the reference signals (i.e., $\{ x_{k,t} \}$ with $t =1$) are unknown and have the same \emph{a priori} distribution as the data. This ensures the rotational invariance of message $m_{cxy_{k,t} \rightarrow g_k}(g_{k})$, i.e., $m_{cxy_{k,t} \rightarrow g_k}(g_{k}) = m_{cxy_{k,t} \rightarrow g_k}(g_{k} e^{\text{j} \theta}), \ \forall \theta \in \Omega_{\cal S}$. This property is useful in developing the efficient RIGM approximation in Section IV-B, which facilitates the design of the low-complexity algorithm and the state evolution analysis in Sections IV and V.

\emph{2) Messages passed from $\{g_k\}$ to $\{cxy_{k,t}\}$:} For the same channel coefficient $g_k$, we have $T$ messages $m_{cxy_{k,t} \rightarrow g_k}(g_{k}), \ t = 1,...,T$. From the sum-product rule, we obtain

\begin{equation}\label{eq.hcom}
\begin{split}
&\hspace{-0.3cm}m_{g_{k},t }(g_{k})  \\
&\hspace{-0.3cm}= \prod_{t^\prime = 1, t^\prime \neq t}^{T} m_{cxy_{k,t'} \rightarrow g_k}(g_{k}) \\
                     &\hspace{-0.3cm}= \prod_{t^\prime = 1, t^\prime \neq t}^{T} \sum_{j=1}^{\vert {\cal S} \vert} \frac{p_{\cal S}(s_j)}{\pi v_{r_{k,t^\prime}}  / \Vert s_j \Vert^2 } {\rm exp}\left(-\frac{\Vert g_{k} - \hat{r}_{k,t^\prime} /s_j \Vert^2}{v_{r_{k,t^\prime}}  / \Vert s_j \Vert^2}\right), \ \forall t
\end{split}
\end{equation}
and
\begin{equation}\label{eq.hcom2}
m_{g_{k} \rightarrow cxy_{k,t}}(g_{k}) = m_{g_{k},t }(g_{k}) p_{G_k}(g_k), \ \forall t,
\end{equation}
where $p_{G_k}(g_k)$ is the \emph{a priori} distribution of the effective channel $g_k$ given in (\ref{eq.pg}).

\emph{3) Messages passed from $\{cxy_{k,t}\}$ to $\{y_{k,t}\}$:} With the message $m_{g_{k} \rightarrow cxy_{k,t}}(g_{k})$ from node $g_k$, we obtain
\begin{equation}\label{eq.h2y}
\begin{split}
& m_{cxy_{k,t} \rightarrow y_{k,t}}(y_{k,t}) \\
& = \int_{g_{k},x_{k,t}} m_{g_{k} \rightarrow cxy_{k,t}}(g_{k}) p_{\cal S}(x_{k,t}) \delta(y_{k,t}-g_k x_{k,t}) \\
&= \sum_{j=1}^{\vert {\cal S} \vert} p_{\cal S}(s_j) m_{g_{k} \rightarrow cxy_{k,t}}(y_{k,t} / s_j), \ \forall t.
\end{split}
\end{equation}
These messages are passed to the variable nodes $\{y_{k,t}\}$ to activate the next SMD operation. The SMD and CSCE operations iterate until convergence.
\subsection{Overall Algorithm and Complexity Analysis}
Based on the procedures described in Sections III-B and III-C, we outline the main steps of the proposed JUICESD algorithm in Algorithm \ref{alg.Juicesd}, where $Q$ is the maximum number of iterations allowed in the SMD module, and $Q'$ is the maximum number of iterations between the SMD and CSCE modules.
\begin{algorithm}[!t]
\caption{JUICESD Algorithm}
\label{alg.Juicesd}
%\State
{\bf Input:} Received signal $\bm{R}$, signature matrix $\bm{A}$, signal distribution $p_{\cal S}(s)$, channel distributions $\{ p_{H_k}(\cdot) \}$, user activity probability $\lambda$.

{\bf for}  $q'=1,2,...,Q'$

\ \ \ {// SMD module}

\ \ \ {\bf Initialization:} $\hat{y}_{k,t}={\rm E}\{y_{k,t}\}$, $v_{y_{k,t}}={\rm Var}\{y_{k,t}\}$, $\hat{s}_{l,t}^{(0)}=0$;

\ \ \ {\bf for}  $q=1,2,...,Q$

\ \  \ \  \  \  Calculate $v_{z_{l,t}}$ and $\hat{z}_{l,t}, \ \forall l,t$, based on (\ref{SMD_P_Variance})-(\ref{SMD_Z_Mean});

\ \  \  \ \  \  Calculate $v_{y_{k,t}}$ and $\hat{y}_{k,t}, \ \forall k,t$, based on (\ref{SMD_S_Variance})-(\ref{SMD_Y_Mean});

\ \ \ {\bf end}

\vspace{0.1cm}
\ \ \ {// CSCE module}

\ \ \ Calculate the message of $g_{k}$ in each time slot using (\ref{eq.y2ht});

\ \ \ Calculate the combined message of $g_{k}$ using (\ref{eq.hcom}) and (\ref{eq.hcom2});

\ \ \ Calculate the refined messages of $\{y_{k,t}\}$ using (\ref{eq.h2y});

{\bf end}

{\bf Output:} Obtain the final estimates $\{ \hat{u}_k \}$, $\{ \hat{h}_k \}$ and $\{ \hat {x}_{k,t} \}$ according to (\ref{eq.ua})-(\ref{eq.sd}).
\end{algorithm}

The complexity of the JUICESD algorithm is described as follows. The algorithm consists of two modules: SMD and CSCE. The SMD operation is based on slot-wise GAMP; hence, its complexity is $\mathcal{O}(KLT)$. The complexity of the CSCE operation is dominated by (\ref{eq.hcom}). Recall that (\ref{eq.hcom}) is the product of $(T-1)$ $|{\cal S}|$-component Gaussian mixtures. A direct evaluation of (\ref{eq.hcom}) results in a complexity of $\mathcal{O}( KT \vert {\cal S} \vert^{T-1} )$ for the CSCE. Consequently, the total complexity of JUICESD is $\mathcal{O}( KLT + KT \vert {\cal S} \vert^{T-1})$.

This complexity is linear in the number of users and the spreading length, but exponential in the frame length $T$. Clearly, the proposed JUICESD algorithm is well suited for the case of massive connections with very short packets. For mMTC scenarios with relatively long packets, its complexity could become unaffordable in practice. To address this issue, we further propose the JUICESD-RIGM algorithm as detailed in Section IV.

Algorithm \ref{alg.Juicesd} assumes that the reference signals are unknown during the iteration; see (\ref{eq.y2ht}) and the discussions therein. Yet, the knowledge of the reference signals $\{x_{k,1}, \forall k \}$ are used only in the final step to remove the phase ambiguity of the output. Alternatively, we may incorporate the knowledge of the reference signals into the iterative process by letting $p_{\cal S}(x_{k,1})=\delta(x_{k,1}-s_p), \ \forall k$, in (\ref{eq.y2ht}). Numerical results show that these two approaches in fact result in almost the same performance. However, a disadvantage for the latter approach is that the rotational invariance property of messages no longer holds in (\ref{eq.y2ht}) and (\ref{eq.hcom}). This would prevent further development of the low-complexity yet efficient algorithm as the rotational invariance property plays an essential role in establishing the next JUICESD-RIGM algorithm and its state evolution analysis.

\section{JUICESD-RIGM Algorithm}\label{sec.JUICESD-CAGMA}
Given that the modulation constellation ${\cal S}$ has a rotational invariance property with respect to the angle set $\Omega_{\cal S}$, we can introduce a rotationally invariant Gaussian mixture (RIGM) model to alleviate the complexity burden of the JUICESD algorithm.
\subsection{Preliminaries}
Recall that (\ref{eq.hcom}) is the product of $(T-1)$ $|{\cal S}|$-component Gaussian mixtures. There are $\vert {\cal S} \vert^{T-1}$ Gaussian components in (\ref{eq.hcom}) in general, and a direct evaluation of (\ref{eq.hcom}) can be very complicated especially for a large $T$. The complexity can be reduced to be linear in $T$ if we discretize each $\vert {\cal S} \vert$-component Gaussian mixture by sampling and then evaluate (\ref{eq.hcom}) approximately based on the discrete samples. However, due to the high dynamic range of the means and variances involved in (\ref{eq.hcom}), the complexity of this sampling approach is still high since a large number of samples are required to ensure a good approximation of (\ref{eq.hcom}).

Another approach is to rely on the Gaussian approximation that is widely used in the design of approximate message passing algorithms \cite{Zhang2018Blind,Ding2018Sparsity,Fan2017Message}. However, simply approximating (\ref{eq.hcom}) by a Gaussian distribution can lead to a substantial information loss and consequently incurs evident performance degradation, as demonstrated by the numerical results in Section \ref{sec.Numerical}.\footnote{For Gaussian approximation, the knowledge of the reference signals should be utilized in the iterative process of the algorithm, i.e., the \emph{a priori} distribution of each $x_{k,1}$ should be set to $p_{\cal S}(x_{k,1})=\delta(x_{k,1}-s_p), \ \forall k$. Otherwise, the algorithm with Gaussian approximation fails by noting that the means of the messages in (\ref{eq.y2ht}) and (\ref{eq.hcom}) are always zeros due to the rotational invariance property.} The reason for this performance degradation is possibly the phase ambiguity aforementioned for problem (\ref{eq.q1}). Due to the phase ambiguity, there are usually multiple Gaussian components with equal importance in (\ref{eq.hcom}), while the number of equally important Gaussian components is determined by the geometric symmetry of the modulation constellation. This inspires us to use a RIGM model for message approximation, as described next.
\subsection{Rotationally Invariant Gaussian Mixture Approximation}
Without loss of generality, we focus on the case of $t = 1$ with $m_{g_{k},t }(g_{k})$ in (\ref{eq.hcom}) given by
\begin{equation}\label{eq.hcom_t1}
\begin{split}
m_{g_{k},1}(g_{k})  &= \prod_{t^\prime = 2}^{T} m_{cxy_{k,t'} \rightarrow g_k}(g_{k}) \\
                    & = \prod_{t^\prime = 2}^{T} \sum_{j=1}^{\vert {\cal S} \vert} \frac{p_{\cal S}(s_j)}{\pi v_{r_{k,t^\prime}}  / \Vert s_j \Vert^2 } {\rm exp}\left(-\frac{\Vert g_{k} - \hat{r}_{k,t^\prime} /s_j \Vert^2}{v_{r_{k,t^\prime}}  / \Vert s_j \Vert^2}\right).
\end{split}
\end{equation}
To develop an efficient approximation method, we rewrite (\ref{eq.hcom_t1}) as
\begin{equation}\label{eq.hcom_gm}
\begin{split}
&m_{g_{k},1}(g_{k}) \\ & = \sum_{j_2, \cdots, j_T = 1}^{\vert {\cal S} \vert} \prod_{t^\prime = 2}^{T}  \frac{p_{\cal S}(s_{j_{t^\prime}})}{\pi v_{r_{k,t^\prime}}  / \Vert s_{j_{t^\prime}} \Vert^2 } {\rm exp}\left(-\frac{\Vert g_{k} - \hat{r}_{k,t^\prime} /s_{j_{t^\prime}} \Vert^2}{v_{r_{k,t^\prime}}  / \Vert s_{j_{t^\prime}} \Vert^2}\right) \\
                    & = \sum_{j_2, \cdots, j_T = 1}^{\vert {\cal S} \vert}   w_{j_2, \cdots, j_T} \mathcal{CN} \left(  \mu_{j_2, \cdots, j_T},\tau_{j_2, \cdots, j_T} \right)
\end{split}
\end{equation}
where $\mu_{j_2, \cdots, j_T}$ and $\tau_{j_2, \cdots, j_T}$ can be obtained by comparing the expressions in (\ref{eq.hcom_gm}), and
the weight $w_{j_2, \cdots, j_T}$ is given by
\begin{equation}\nonumber
w_{j_2, \cdots, j_T}  = \frac{ \prod\nolimits_{t^\prime = 2}^{T} \frac{p_{\cal S}(s_{j_{t^\prime}})}{\pi v_{r_{k,t^\prime}}  / \Vert s_{j_{t^\prime}} \Vert^2 } {\rm exp}\left(-\frac{\Vert g_{k} - \hat{r}_{k,t^\prime} /s_{j_{t^\prime}} \Vert^2}{v_{r_{k,t^\prime}}  / \Vert s_{j_{t^\prime}} \Vert^2}\right) } {\mathcal{CN} \left(  \mu_{j_2, \cdots, j_T},\tau_{j_2, \cdots, j_T} \right)}.
\end{equation}

\begin{figure}
\centering
\subfigure[QPSK symbols]{
\label{Fig.QPSKcombine1}
\includegraphics[scale=0.5]{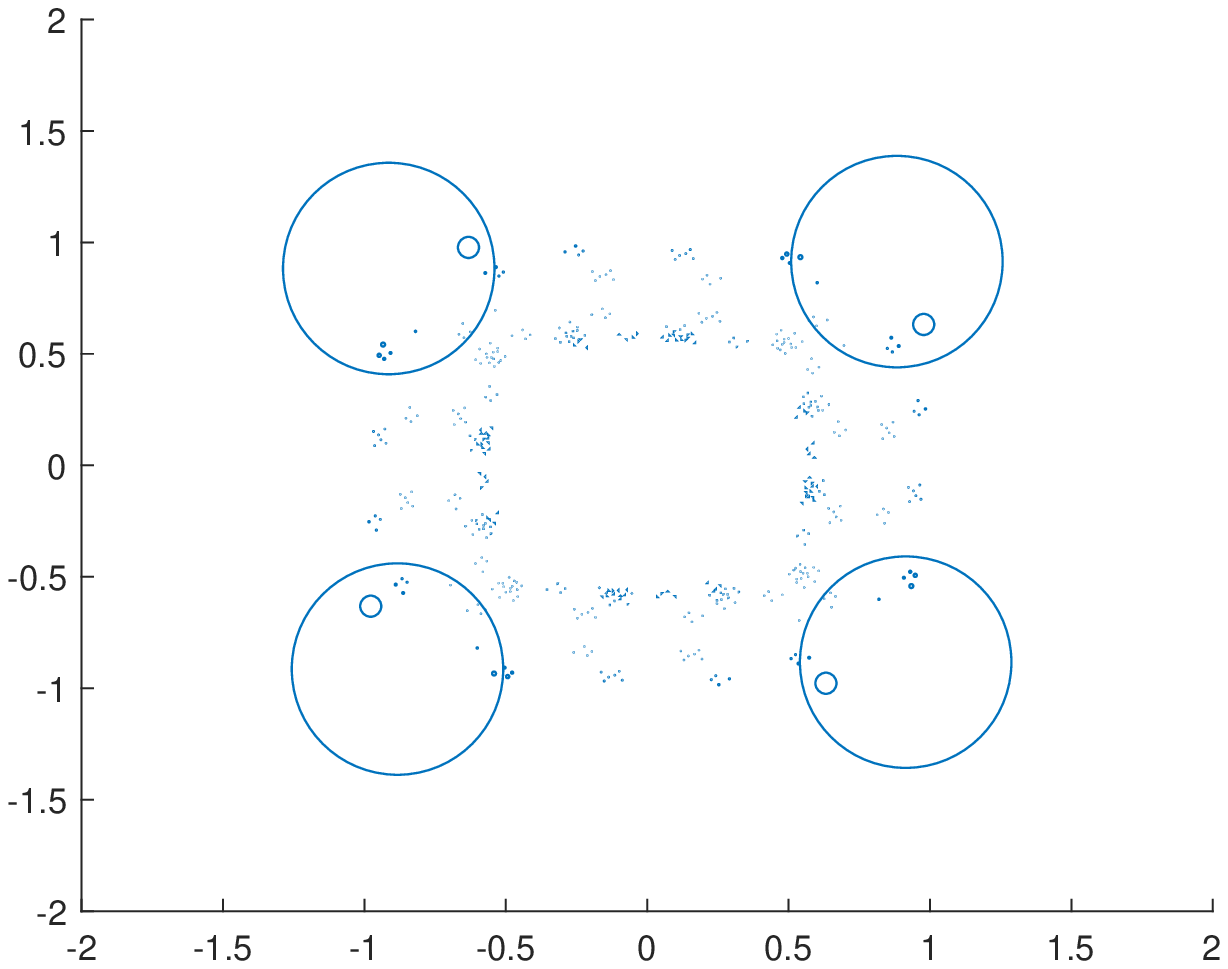}}
\vspace{0cm}
\centering
\subfigure[16QAM symbols]{
\label{Fig.16QAMcombine1}
\includegraphics[scale=0.5]{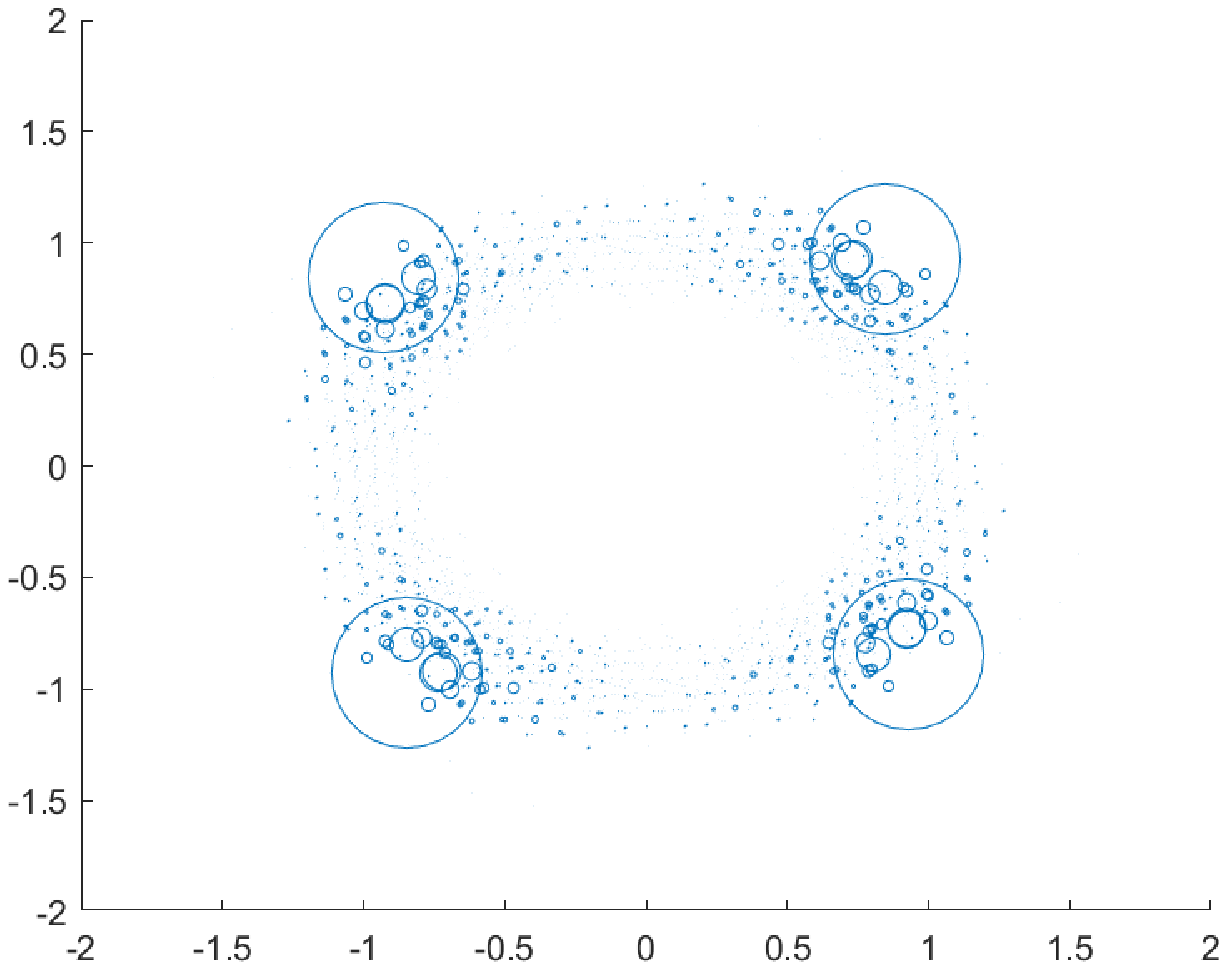}}
\caption{An illustration of the Gaussian components in (\ref{eq.hcom_gm}) for (a) QPSK and (b) 16QAM.}
\label{Fig.GMcombine1}
\end{figure}

In (\ref{eq.hcom_gm}), there are $\vert {\cal S} \vert^{T-1}$ Gaussian components with different weights, each of which corresponds to one possible realization of the transmit symbol vector $[s_{j_2}, \cdots, s_{j_T}]$. Figs. \ref{Fig.QPSKcombine1} and \ref{Fig.16QAMcombine1} show some typical examples of the Gaussian components in (\ref{eq.hcom_gm}) for QPSK and 16QAM, respectively, where each circle represents one Gaussian component with the center for its mean value and the radius for its weight. From Fig. \ref{Fig.GMcombine1}, we clearly see that the GM distribution in (\ref{eq.hcom_gm}) is rotationally invariant with respect to the set $\Omega_{\cal S}$, i.e., any Gaussian component in (\ref{eq.hcom_gm}) rotated by an angle in $\Omega_{\cal S}$ is still a Gaussian component in (\ref{eq.hcom_gm}). From Fig. \ref{Fig.GMcombine1}, we also see that the GM distribution in (\ref{eq.hcom_gm}) is usually dominant by $\vert \Omega_{\cal S} \vert$ equally important Gaussian components. This inspires us to approximate (\ref{eq.hcom}) by the model below:

\begin{equation}\label{eq.fourGaussApproximation}
\hat{m}_{g_{k},t }(g_{k})=\sum_{i=1}^{|\Omega_{\cal S}|} \frac{1}{|\Omega_{\cal S}| \pi v_{g_{k},t}}{\rm exp}\left( -\frac{\Vert g_k-\hat{g}_{k,t,i} \Vert^2}{v_{g_{k},t}} \right), %\ \forall k,t
\end{equation}
where $\hat{g}_{k,t,i}$ is the mean of the $i$-th Gaussian component satisfying $\hat{g}_{k,t,i} = e^{\text{j} (i-1) \theta_0}  \hat{g}_{k,t,1}$, and $v_{g_{k},t}$ is the common variance for all the Gaussian components. We note that the GM in (\ref{eq.fourGaussApproximation}) is rotationally invariant over $\Omega_{\cal S}$. This rotational invariance property enables the characterization of (\ref{eq.fourGaussApproximation}) by only two parameters, namely, the component mean $\hat{g}_{k,t,i}$ and the variance $v_{g_{k},t}$.

We now discuss how to determine the parameters $\{\hat{g}_{k,t,i}\}$ and $v_{g_{k},t}$ in (\ref{eq.fourGaussApproximation}). We may obtain them by minimizing the KL divergence between $m_{g_{k},t }(g_{k})$ in (\ref{eq.hcom}) and $\hat{m}_{g_{k},t }(g_{k})$ in (\ref{eq.fourGaussApproximation}), i.e.,
\begin{equation}\label{eq.fourGaussApproximation_KL}
[\{\hat{g}_{k,t,i}\}, v_{g_{k},t}] = \mathop {\arg \min }\limits_{ \{\hat{g}_{k,t,i}\}, v_{g_{k},t}  } D_{KL}\left( m_{g_{k},t }(g_{k}) \parallel \hat{m}_{g_{k},t }(g_{k}) \right).
\end{equation}
However, due to the large number of Gaussian components involved in $m_{g_k,t}(g_{k})$, the calculation in (\ref{eq.fourGaussApproximation_KL}) is generally complicated.

We next describe a recursive approximate method to calculate the parameters $\{\hat{g}_{k,t,i}\}$ and $\{v_{g_{k},t}\}$ in (\ref{eq.fourGaussApproximation}). For brevity, we focus on the case of $t = 1$, i.e., the approximation of $m_{g_{k},1}(g_{k})$ in (\ref{eq.hcom_t1}).
We adopt a recursive method as follows. Assume that the first $T^\prime-1$ factors in the product in (\ref{eq.hcom_t1}) have been approximated by a $\vert \Omega_{\cal S} \vert$-component RIGM, denoted by $m_{g_{k},1}^{(T^\prime)}(g_{k})$. We then approximate the product of $m_{g_{k},1}^{(T^\prime)}(g_{k})$ and $m_{cxy_{k,T^\prime+1} \rightarrow g_k}(g_{k})$ by a $\vert \Omega_{\cal S} \vert$-component RIGM, denoted by $m_{g_{k},1}^{(T^\prime+1)}(g_{k})$. This process continues until $m_{g_{k},1}^{(T)}(g_{k})$ is obtained, which gives the approximation of (\ref{eq.hcom_t1}) in the form of (\ref{eq.fourGaussApproximation}). Below we present the details involved in the recursive method.

\emph{1) Initialization:} We need to approximate $m_{cxy_{k,2} \rightarrow g_k}(g_{k})$ by a $\vert \Omega_{\cal S} \vert$-component RIGM.
Recall that ${\cal S}$ is rotationally invariant with respect to $\Omega_{\cal S} = \{\theta_0, 2\theta_0, \cdots, \vert \Omega_{\cal S} \vert\theta_0 \}$. We divide ${\cal S}$ into $\vert \Omega_{\cal S} \vert$ subsets, each with $\vert {\cal S} \vert / \vert \Omega_{\cal S} \vert$ elements. Each subset $i$ contains the constellation points with the phase angles falling into the range of $\left[ (i-1)\theta_0,i\theta_0 \right)$, for $i=1,\cdots,\vert \Omega_{\cal S} \vert$. We denote the $i$-th subset as
\begin{equation}\label{eq.subset}
{\cal S}_i \doteq \{ s_{i,1}, s_{i,2}, \cdots, s_{i,{\vert {\cal S} \vert / \vert \Omega_{\cal S} \vert}} \}, \ i = 1, 2, \cdots, \vert \Omega_{\cal S} \vert.
\end{equation}
It can be verified that $\{ {\cal S}_i \}$ is a partition of ${\cal S}$, and $ {\cal S}_{i}  = e^{ \text{j} (i-1)\theta_0} {\cal S}_1$.

Using (\ref{eq.subset}), we rewrite $m_{cxy_{k,2} \rightarrow g_k}(g_{k})$ in (\ref{eq.hcom_t1}) (the multiplicative component for $t^\prime = 2$) as
\begin{equation}\label{eq.y2ht_subset}
\sum_{i=1}^{\vert \Omega_{\cal S} \vert} \sum_{m=1}^{\vert {\cal S} \vert / \vert \Omega_{\cal S} \vert} \frac{ 1 }{ {\vert {\cal S} \vert} \pi v_{r_{k,2}}  / \Vert s_{i,m} \Vert^2 } {\rm exp}\left(-\frac{\Vert g_{k} - \hat{r}_{k,2} /s_{i,m} \Vert^2}{ v_{r_{k,2}}  / \Vert s_{i,m} \Vert^2}\right).
\end{equation}
We then approximate the ${\vert {\cal S} \vert / \vert \Omega_{\cal S} \vert}$ Gaussian components in (\ref{eq.y2ht_subset}) related to subset ${\cal S}_i$ by a single Gaussian component and obtain the $\vert \Omega_{\cal S} \vert$-component GM approximation of (\ref{eq.y2ht_subset}) as
\begin{equation}\label{eq.GAM}
\sum_{i=1}^{\vert \Omega_{\cal S} \vert} \frac{1}{|\Omega_{\cal S}| \pi v_{g_k,i}^{(2)}}{\rm exp}\left( -\frac{\Vert g_k -\hat{g}_{k,i}^{(2)} \Vert^2}{v_{g_k,i}^{(2)}} \right), %\ \forall k,t
\end{equation}
where
\begin{subequations}\label{eq.MeansAndVariances}
\begin{equation}
\hat{g}_{k,i}^{(2)} = \frac{1}{\vert {\cal S} \vert/ \vert \Omega_{\cal S} \vert} \sum _{m=1}^{\vert {\cal S} \vert/ \vert \Omega_{\cal S} \vert} \frac{\hat{r}_{k,2} }{s_{i,m}},
\end{equation}
\begin{equation}
v_{g_k,i}^{(2)} = \frac{1}{\vert {\cal S} \vert/ \vert \Omega_{\cal S} \vert} \sum_{m=1}^{\vert {\cal S} \vert/ \vert \Omega_{\cal S} \vert} \left( \frac{v_{r_{k,2}} }{\Vert s_{i,m} \Vert^2}+ \left \Vert \frac{\hat{r}_{k,2} }{s_{i,m}} \right \Vert^2 \right)- \Vert \hat{g}_{k,i}^{(2)} \Vert^2.
\end{equation}
\end{subequations}

With $ {\cal S}_{i}  = e^{ \text{j} (i-1)\theta_0} {\cal S}_1$, we can verify in (\ref{eq.GAM}) that: $1) \hat{g}_{k,i}^{(2)} = e^{\text{j} (i-1) \theta_0}  \hat{g}_{k,1}^{(2)}, \forall i$, and $2) \{ v_{g_k,i}^{(2)}, \forall i \}$ share a common value denoted by $v_{g_k}^{(2)}$. Hence, (\ref{eq.GAM}) satisfies the RIGM requirement.

\begin{figure}
\centering
\subfigure[Approximating a $\vert \cal S \vert$-component GM by a $\vert \Omega_{\cal S} \vert$-component GM]{
\label{fig:Approximation_CAGMA1}
\includegraphics[scale=0.35]{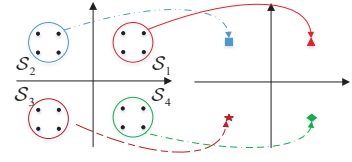}}
\vspace{0cm}
\centering
\subfigure[Approximating the product of a ${\vert \Omega_{\cal S} \vert}$-component GM and a ${\vert \cal S \vert}$-component GM by a new $\vert \Omega_{\cal S} \vert$-component GM]{
\label{fig:Approximation_CAGMA2}
\includegraphics[scale=0.35]{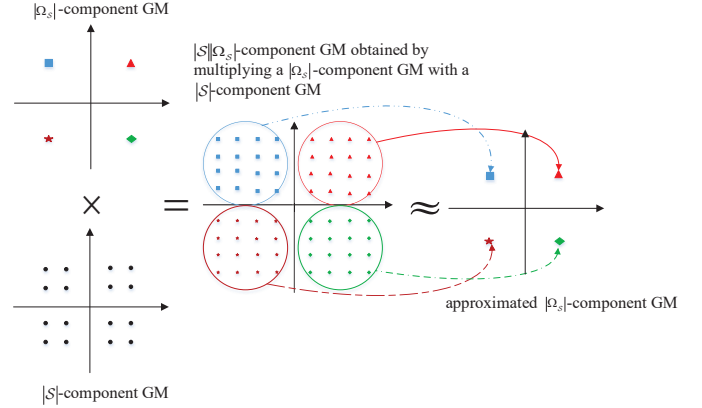}}
\caption{An illustration of the basic operations in approximating (\ref{eq.hcom}) by (\ref{eq.fourGaussApproximation}).}
\label{Fig.Approximation_CAGMA}
\end{figure}

Fig. \ref{fig:Approximation_CAGMA1} shows an example of the approximation process in (\ref{eq.subset})-(\ref{eq.GAM}) for the 16QAM constellation, with $\vert {\cal S} \vert= 16$, $\vert \Omega_{\cal S} \vert = 4$, and $\theta_0 = 0.5 \pi$. In Fig. \ref{fig:Approximation_CAGMA1}, the left subplot represents the $\vert {\cal S} \vert$-component GM in (\ref{eq.y2ht_subset}), with each point corresponding to one Gaussian component related to the constellation point $s_{i,m}, i \in \{1, \cdots, \vert \Omega_{\cal S} \vert \}, m \in \{1, \cdots, \vert {\cal S} \vert / \vert \Omega_{\cal S} \vert \}$. Using the partition $\{ {\cal S}_i, i = 1, \cdots, \vert \Omega_{\cal S} \vert \}$ in (\ref{eq.subset}), these components are divided into $\vert \Omega_{\cal S} \vert$ subsets. Then for each $i \in \{1, \cdots, \vert \Omega_{\cal S} \vert\}$, all Gaussian components in (\ref{eq.y2ht_subset}) related to the constellation points in ${\cal S}_i$ are approximated by one Gaussian distribution, as shown by the arrow in Fig. \ref{fig:Approximation_CAGMA1}. This results in the $\vert \Omega_{\cal S} \vert$-component GM approximation in (\ref{eq.GAM}) shown in the right subplot in Fig. \ref{fig:Approximation_CAGMA1}.

\emph{2) Recursion:} We then need to approximate the product of $m_{g_{k},1}^{(T^\prime)}(g_{k})$ and $m_{cxy_{k,T^\prime+1} \rightarrow g_k}(g_{k})$ by a $\vert \Omega_{\cal S} \vert$-component RIGM.
To this end, we multiply each Gaussian component in $m_{g_{k},1}^{(T^\prime)}(g_{k})$ with $m_{cxy_{k,T^\prime+1} \rightarrow g_k}(g_{k})$, and then approximate the result by a new Gaussian component.

Suppose that $m_{g_{k},1}^{(T^\prime)}(g_{k})$ is given by
\begin{multline}\label{eq.fourGaussApproximation_temp}
m_{g_{k},1}^{(T^\prime)}(g_{k}) = \sum_{i=1}^{|\Omega_{\cal S}|} \frac{1}{|\Omega_{\cal S}| \pi v_{g_{k}}^{(T')}}  {\rm exp}\left( -\frac{\Vert g_k-\hat{g}_{k,i}^{(T')} \Vert^2}{v_{g_{k}}^{(T')}} \right), \\  \text{for} \ T'=2,\cdots,T-1,
\end{multline}
where $\hat{g}_{k,i}^{(T')}$ is the mean of the $i$-th Gaussian component satisfying $\hat{g}_{k,i}^{(T')} = e^{\text{j} (i-1) \theta_0}  \hat{g}_{k,1}^{(T')}$, and $v_{g_{k}}^{(T')}$ is the common variance for all the Gaussian components.
For $T'=2$, (\ref{eq.fourGaussApproximation_temp}) is initialized by (\ref{eq.GAM}).

The product of $m_{cxy_{k,T^\prime+1} \rightarrow g_k}(g_{k})$ and the $i$-th Gaussian component of $m_{g_{k},1}^{(T^\prime)}(g_{k})$ is still a GM, expressed by $\sum \nolimits_{j=1}^{\vert {\cal S} \vert} w_{i,j}^{(T')} \mathcal{CN} (\mu_{i,j}^{(T')}, \tau_{i,j}^{(T')})$, where

\begin{subequations}\label{eq.WeightsMeansVariances}
\begin{equation}\label{weights_multi_Gaussian}
w_{i,j}^{(T')}  \propto \frac{{\rm exp}\left( - \frac{\Vert \hat{g}_{k,i}^{(T')}-\hat{r}_{k,T^\prime+1}  /  s_j \Vert^2 }{ v_{g_{k}}^{(T')} + v_{r_{k,T^\prime+1}}  / \Vert s_j \Vert^2}  \right)}{\pi ( v_{g_{k}}^{(T')} + v_{r_{k,T^\prime+1}}  / \Vert s_j \Vert^2)} \ \text{with} \ \sum_{j=1}^{\vert {\cal S} \vert} w_{i,j}^{(T')} = 1,
\end{equation}

\begin{equation}\label{means_variances_multi_Gaussian}
\begin{split}
\mu_{i,j}^{(T')} = \frac{ \hat{g}_{k,i}^{(T')} v_{r_{k,T^\prime+1}}  / \Vert s_j \Vert^2 + v_{g_{k}}^{(T')} \hat{r}_{k,T^\prime+1}  / s_j  }{ v_{g_{k}}^{(T')} + v_{r_{k,T^\prime+1}}  / \Vert s_j \Vert^2 },
\end{split}
\end{equation}
\begin{equation}
\hspace{-3.1cm}\tau_{i,j}^{(T')} = \frac{v_{g_{k}}^{(T')} v_{r_{k,T^\prime+1}}  / \Vert s_j \Vert^2}{v_{g_{k}}^{(T')} + v_{r_{k,T^\prime+1}}  / \Vert s_j \Vert^2}.
\end{equation}
\end{subequations}
By using the moment matching principle, we approximate this GM by a new Gaussian distribution $\mathcal{CN}(\hat{g}_{k,i}^{(T'+1)},v_{g_{k},i}^{(T'+1)})$ with
\begin{subequations}\label{combined_mean_variance_multi_Gaussian}
\begin{equation}
\hat{g}_{k,i}^{(T'+1)} = \sum_{j=1}^{\vert {\cal S} \vert} w_{i,j}^{(T')}\mu_{i,j}^{(T')},
\end{equation}
\begin{equation}
v_{g_{k},i}^{(T'+1)}  =\sum_{j=1}^{\vert {\cal S} \vert} w_{i,j}^{(T')} ( \tau_{i,j}^{(T')} + \Vert \mu_{i,j}^{(T')} \Vert^2 ) -\Vert \hat{g}_{k,i}^{(T'+1)} \Vert^2.
\end{equation}
\end{subequations}

Performing the same operation for the other ${\vert \Omega_{\cal S} \vert} - 1$ components in $m_{g_{k},1}^{(T^\prime)}(g_{k})$, we obtain a $\vert \Omega_{\cal S} \vert$-component GM
\begin{equation}\label{eq.fourGaussApproximation_temp2}
m_{g_{k},1}^{(T^\prime+1)}(g_{k}) = \sum_{i=1}^{|\Omega_{\cal S}|} \frac{{\rm exp}\left( -\frac{\Vert g_k-\hat{g}_{k,i}^{(T'+1)} \Vert^2}{v_{g_{k}}^{(T'+1)}} \right)}{|\Omega_{\cal S}| \pi v_{g_{k}}^{(T'+1)}}.
\end{equation}

Fig. \ref{fig:Approximation_CAGMA2} illustrates the operations in (\ref{eq.fourGaussApproximation_temp})-(\ref{eq.fourGaussApproximation_temp2}). The upper left subplot represents the $\vert \Omega_{\cal S} \vert$-component GM in (\ref{eq.fourGaussApproximation_temp}) while the lower left subplot represents the $\vert {\cal S} \vert$-component GM message $m_{cxy_{k,T^\prime+1} \rightarrow g_k}(g_{k})$. Multiplying each Gaussian component in the upper left GM with the lower left GM results in one $\vert {\cal S} \vert$-component GM in a circle in the middle subplot, with parameters defined in (\ref{eq.WeightsMeansVariances}). Then the $\vert {\cal S} \vert$-component GM in each circle in the middle subplot is approximated by one Gaussian distribution, as shown by the arrow in the figure, with parameters calculated by (\ref{combined_mean_variance_multi_Gaussian}). This approximation operation gives the $\vert \Omega_{\cal S} \vert$-component GM in (\ref{eq.fourGaussApproximation_temp2}) shown in the right subplot in Fig. \ref{fig:Approximation_CAGMA2}.

The rotational invariance property of (\ref{eq.fourGaussApproximation_temp2}) is ensured by the following lemma.

\begin{lemma}\label{Lemma.1}
The message in (\ref{eq.fourGaussApproximation_temp2}) is rotationally invariant over the angle set $\Omega_{\cal S}$, i.e.,
\begin{subequations}\label{eq.reduced}
\begin{equation} \label{eq.reducedmean}
\hat{g}_{k,i}^{(T'+1)} = e^{\text{j} (i-1) \theta_0} \hat{g}_{k,1}^{(T'+1)},\ \forall i,
\end{equation}
\begin{multline}\label{eq.reducedvar}
v_{g_{k},1}^{(T'+1)} =  \cdots = v_{g_{k},|\Omega_{\cal S}|}^{(T'+1)} \equiv v_{g_{k}}^{(T'+1)}, \ 2 \leq T'\leq T-1.
\end{multline}
\end{subequations}
\end{lemma}
\emph{Proof}: See Appendix A.

\subsection{CSCE Based on RIGM Approximation}
With the replacement of (\ref{eq.hcom}) by (\ref{eq.fourGaussApproximation}), we can rewrite (\ref{eq.h2y}) as
\begin{multline}\label{eq.BernoullifourGaussApproximationFinal}
 m_{cxy_{k,t} \rightarrow y_{k,t}}(y_{k,t}) = w_{k,t} \delta(y_{k,t}) \\ +  \sum_{j=1}^{\vert {\cal S} \vert} \frac{1-w_{k,t}}{\vert {\cal S} \vert \pi v_{y_{k,t,j}}}{\rm exp} \left( -\frac{ \Vert y_{k,t} - \hat{y}_{k,t,j} \Vert^2 }{v_{y_{k,t,j}}} \right)
\end{multline}
with the weight of the impulse term given by
\begin{subequations}\label{eq:BernoullifourGaussApproximationFinal_Parameters}
\begin{equation}\label{eq.BernoullifourGaussApproximationFinal_Weight}
w_{k,t} = \frac{1}{1+\tilde{w}_{k,t}}
\end{equation}
where
\begin{equation}
\tilde{w}_{k,t} = \frac{\sum \nolimits_{i=1}^{\vert \Omega_{\cal S} \vert} \frac{\lambda}{{|\Omega_{\cal S}|} \pi (v_{g_{k},t}+\beta_k)} {\rm exp} \left( -\frac{\Vert \hat{g}_{k,t,i} \Vert^2}{v_{g_{k},t}+\beta_k} \right)}{\sum \nolimits_{i=1}^{|\Omega_{\cal S}|} \frac{1-\lambda}{{|\Omega_{\cal S}|} \pi v_{g_{k},t}} {\rm exp} \left( -\frac{\Vert \hat{g}_{k,t,i} \Vert^2}{v_{g_{k},t}} \right)} ,
\end{equation}
and the variance and the mean of the $j$-th Gaussian component are respectively
\begin{equation} \label{eq.BernoullifourGaussApproximationFinal_Variance}
v_{y_{k,t,j}} = \frac{\beta_k v_{g_{k},t} \Vert s_j \Vert^2}{\beta_k+v_{g_{k},t}} \ \text{and} \ \hat{y}_{k,t,j} = \frac{\hat{g}_{k,t,1} \beta_k s_j}{\beta_k + v_{g_{k},t}}.
\end{equation}
\end{subequations}
Finally, the JUICESD-RIGM algorithm is obtained by replacing (\ref{eq.hcom}) and (\ref{eq.h2y}) in the JUICESD algorithm with (\ref{eq.fourGaussApproximation}) and (\ref{eq.BernoullifourGaussApproximationFinal}), respectively.

\subsection{Overall Algorithm}
\begin{algorithm}[!t]
\caption{JUICESD-RIGM Algorithm}
\label{alg.Juicesd-CAGMA}
%\State
{\bf Input:} Received signal $\bm{R}$, signature matrix $\bm{A}$, signal distribution $p_{\cal S}(s)$, channel distributions $\{ p_{H_k}(\cdot) \}$, user activity probability $\lambda$.

{\bf for}  $q'=1,2,...,Q'$

\ \ \ {// SMD module}

\ \ \ {\bf Initialization:} $\hat{y}_{k,t}={\rm E}\{y_{k,t}\}$, $v_{y_{k,t}}={\rm Var}\{y_{k,t}\}$, $\hat{s}_{l,t}^{(0)}=0$;

\ \ \ {\bf for}  $q=1,2,...,Q$

\ \  \ \  \  \  Calculate $v_{z_{l,t}}$ and $\hat{z}_{l,t}, \ \forall l,t$, based on (\ref{SMD_P_Variance})-(\ref{SMD_Z_Mean});

\ \  \  \ \  \  Calculate $v_{y_{k,t}}$ and $\hat{y}_{k,t}, \ \forall k,t$, based on (\ref{SMD_S_Variance})-(\ref{SMD_Y_Mean});

\ \ \ {\bf end}

\ \ \ {// CSCE module}

\ \ \  Calculate the message of $g_{k}$ in each time slot using (\ref{eq.y2ht});

\ \ \  Calculate the combined message of $g_{k}$ using (\ref{eq.fourGaussApproximation}) and (\ref{eq.hcom2});

\ \ \  Calculate the refined messages of $\{y_{k,t}\}$ using (\ref{eq.BernoullifourGaussApproximationFinal});

{\bf end}

{\bf Output:} Obtain the final estimates $\{ \hat{u}_k \}$, $\{ \hat{h}_k \}$ and $\{ \hat {x}_{k,t} \}$ according to (\ref{eq.ua})-(\ref{eq.sd}).
\end{algorithm}
The proposed JUICESD-RIGM algorithm is outlined in Algorithm \ref{alg.Juicesd-CAGMA}. Based on the rotational invariance property in Lemma \ref{Lemma.1}, the complexity of evaluating (\ref{eq.fourGaussApproximation}) in the CSCE operation is $\mathcal{O}( \vert {\cal S} \vert T)$ (with the complexities of computing (\ref{eq.GAM}) and (\ref{combined_mean_variance_multi_Gaussian}) given by $\mathcal{O}( \vert {\cal S} \vert / \vert \Omega_{\cal S} \vert)$ and $\mathcal{O}( \vert {\cal S} \vert )$, respectively). Following the arguments in the JUICESD case, we see that the total complexity of the proposed JUICESD-RIGM algorithm becomes $\mathcal{O}(KLT + KT^2 \vert {\cal S} \vert)$. This complexity is clearly much lower than that of the original JUICESD algorithm.

\section{State Evolution of JUICESD-RIGM}\label{StateEvolution}
\begin{figure}[!t]
\centering
\subfigure[Structure of the proposed iterative algorithm]{
\label{fig:SimulationFrame}
\includegraphics[scale=1.1]{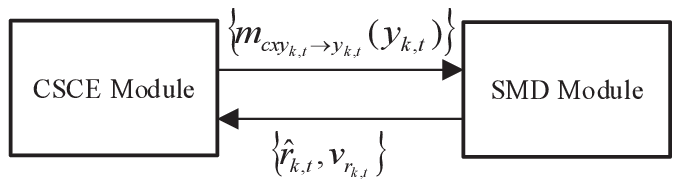}}
\subfigure[The corresponding state evolution]{
\label{fig:StateEvolutionFrame}
\includegraphics[scale=1.1]{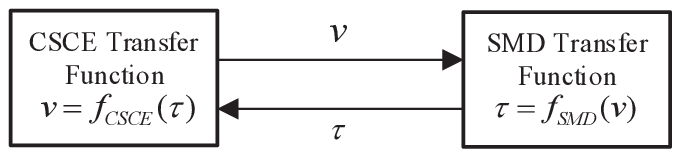}}
\caption{Structure of the proposed iterative algorithm and the corresponding state evolution.}
\label{Fig.Frames}
\end{figure}

We now describe the state evolution of the JUICESD-RIGM algorithm. Recall that in JUICESD-RIGM, we divide the whole receiver into two modules, i.e., the SMD module and the CSCE module as shown in Fig. \ref{fig:SimulationFrame}. Our approach is to characterize the behavior of each module by tracking their input and output mean-square errors (MSEs).

More specifically, as illustrated in Fig. \ref{fig:StateEvolutionFrame}, we aim to characterize the behavior of the SMD module by the transfer function $\tau=f_{SMD}(v)$ where $v$ and $\tau$ are the input and output MSEs of the SMD module, respectively. Correspondingly, the CSCE module can be characterized by the transfer function $v=f_{CSCE}(\tau)$ where $\tau$ and $v$ are the input and output MSEs of the CSCE module, respectively. Then, the performance of the JUICESD-RIGM algorithm is determined by the fixed point of
\begin{equation}\label{eq.OverallStateEvolution}
v = f_{CSCE}(f_{SMD}(v)).
\end{equation}

\subsection{SMD Transfer Function}\label{SSDEvolution}
Consider the establishment of the SMD transfer function $\tau=f_{SMD}(v)$. It is desirable to directly express the generation model of the input messages of the SMD module in (\ref{eq.BernoullifourGaussApproximationFinal}) by the single parameter $v$. Unfortunately, this is difficult since
(\ref{eq.BernoullifourGaussApproximationFinal}) is determined by three sets of variables $\{w_{k,t}\}$, $\{\hat{y}_{k,t,j}\}$, and $\{v_{y_{k,t,j}}\}$. Instead of directly generating the messages in (\ref{eq.BernoullifourGaussApproximationFinal}), we notice that the messages in (\ref{eq.BernoullifourGaussApproximationFinal}) are actually determined by the messages in (\ref{eq.fourGaussApproximation}). We can model the messages in (\ref{eq.fourGaussApproximation}) using a single parameter $v_g$. Specifically, we model each $\hat{g}_{k,t}$ by
\begin{equation}\label{eq.SMD_Evo_hat_gk}
\hat{g}_{k,t} = g_k + \sqrt{v_{g}} \zeta_{k,t}
\end{equation}
where $g_k \sim f_{G_k}(g_k)$,  and each $\zeta_{k,t}$ is independently drawn from $\mathcal{CN}(0,1)$. Then, from the rotational invariance property, the whole set $\{\hat{g}_{k,t,i}\}$ is given by $\{ e^{- \text{j} \theta_i} \hat{g}_{k,t},\ \theta_i \in \Omega_{\cal S} , \ i=1,\cdots, \vert \Omega_{\cal S} \vert  \}$. We next show how to determine $v_g$ for a given $v$. Recall that $\{ v_{y_{k,t,j}}, j = 1,\cdots, {\vert {\cal S} \vert} \}$ in (\ref{eq.BernoullifourGaussApproximationFinal}) are the variances with respect to different component Gaussian distributions. We define the input MSE $v$ to the SMD module as
\begin{equation}\label{eq.SMD_Evolution_V}
v = \frac{1}{KT \vert {\cal S} \vert}\sum_{k, t, j} v_{y_{k,t,j}}.
\end{equation}
Substituting the first equation of (\ref{eq.BernoullifourGaussApproximationFinal_Variance}) into (\ref{eq.SMD_Evolution_V}) and letting $v_{g_{k,t}}=v_g$ for $\forall k,t$, we obtain
\begin{equation}\label{eq.SMD_Evo_V_gk}
%v = \frac{\beta v_{g}}{\vert {\cal S} \vert(\beta + v_{g})} \sum_{j=1}^{\vert {\cal S} \vert} \Vert s_j \Vert^2.
v = \frac{1}{\vert {\cal S} \vert K}\sum_{k} \frac{\beta_k v_g}{\beta_k+v_g} \sum_{j=1}^{\vert {\cal S} \vert} \Vert s_j \Vert^2 = \frac{1}{K}\sum_{k} \frac{\beta_k v_g}{\beta_k+v_g}.
\end{equation}
Then $v_g$ can be obtained according to the given $v$ by solving (\ref{eq.SMD_Evo_V_gk}).

We now summarize the generation process of the input messages. We first calculate $v_g$ from $v$ using (\ref{eq.SMD_Evo_V_gk}), and then generate $\{ \hat{g}_{k,t} \}$ using (\ref{eq.SMD_Evo_hat_gk}). We rotate $\hat{g}_{k,t}$ based on $\Omega_{\cal S}$ to obtain $\{\hat{g}_{k,t,i}\}$ and then the messages in (\ref{eq.fourGaussApproximation}). Finally, we obtain the input messages to SMD in the form of (\ref{eq.BernoullifourGaussApproximationFinal}). With the above input model, we construct the SMD transfer function $\tau = f_{SMD}(v)$ as follows.

Recall that the SMD module is an AMP-type algorithm. Similarly to the state evolution of the AMP algorithm \cite{Bayati2011The}, we derive the SMD transfer function by tracking the equivalent noise-and-interference power seen by each user. Given the input messages $\{ m_{cxy_{k,t} \rightarrow y_{k,t}}(y_{k,t}) \}$, let $Y_{k,t}$ be a random variable with the distribution $ m_{cxy_{k,t} \rightarrow y_{k,t}}(y_{k,t})$. We initialize the equivalent noise-and-interference power seen by each user as
\begin{equation}\label{eq.SSDEvolution0}
\tau^{(0)} \equiv \frac{1}{KT} \sum_{k,t} \left( N_0 + \frac{K-1}{L} E_{ Y_{k,t} } \{|Y_{k,t}|^2\} \right),
\end{equation}
where $E_{ Y_{k,t} } \{|Y_{k,t}|^2\}$ is the interference power caused by $Y_{k,t}$ when there is no interference cancellation; the term $K-1$ implies that each user suffers from the interference of $K-1$ users; and the term $1/L$ comes from the spreading/de-spreading operations. Then we calculate $\tau^{(q)}$ recursively for $q \geqslant 0$ as follows. With $\tau^{(q)}$ given, each user performs the \emph{a posteriori} estimation to obtain a refined estimate with a reduced residual interference power. Let $E \{ Y_{k,t} | Y_{k,t} + \sqrt{\tau^{(q)}} \epsilon, \tau^{(q)} \}$ be the \emph{a posteriori} estimate function in (\ref{SMD_Y_Mean}) in the SMD module, where $Y_{k,t} \sim m_{cxy_{k,t} \rightarrow y_{k,t}}(y_{k,t})$ and $\epsilon \sim \mathcal{CN}(0,1)$. Then the residual interference power of $Y_{k,t}$ after its \emph{a posteriori} estimation is $V_{Y_{k,t}} = E_{\{Y_{k,t}, \epsilon \}} \left( \left | E \{ Y_{k,t} | Y_{k,t} + \sqrt{\tau^{(q)}} \epsilon, \tau^{(q)} \} -Y_{k,t} \right | ^2\right)$.

Consequently, the equivalent noise-and-interference power seen by each user is updated by
\begin{equation}\label{eq.SSDEvolution1}
\tau^{(q+1)} = \frac{1}{KT} \sum_{k,t} \left( N_0 + \frac{K-1}{L} V_{Y_{k,t}} \right).
\end{equation}
$\tau^{(q+1)}$ in (\ref{eq.SSDEvolution1}) is calculated recursively, and the fixed point $\tau$ gives the output MSE of SMD. Then the function $\tau = f_{SMD}(v)$ is obtained.

\vspace{-0.3cm}
\subsection{CSCE Transfer Function}\label{CCEEvolution}
We now describe how to determine the CSCE transfer function $v = f_{CSCE}(\tau)$. As analogous to the SMD case, we need to construct a generation model of the input messages $\{\hat{r}_{k,t} ,v_{r_{k,t}} \}$ of the CSCE by using a single parameter $\tau$. In this regard, we model each $\hat{r}_{k,t} $ as
\begin{equation}\label{eq.CSCE_Evolution_model}
\hat{r}_{k,t} = h_k u_k x_{k,t} +\sqrt{\tau}\zeta_{k,t}, \ \text{for} \ \forall k,t,
\end{equation}
where $h_k \sim p_{H_k}(h_k)$, $x_{k,t} \sim p_{\cal S}(x_{k,t})$, and $\zeta_{k,t}\sim {\mathcal {CN}}(0,1)$.
We observe from numerical results that the output MSEs for the active users and inactive users can be quite different. Note that the user activity state $u_k$ can be estimated by the algorithm with very high accuracy, especially for a relatively large $T$. And for an inactive user with known $u_k = 0$, both its channel estimation error and its symbol error rate can be regarded as zero. In this case, we will only consider the active users with $u_k=1$ to simplify the analysis.

With the input messages $\{\hat{r}_{k,t} \}$ generated based on (\ref{eq.CSCE_Evolution_model}) and assuming $v_{r_{k,t}} = \tau,\ \forall k,t$, we obtain the messages of the effective channels $\hat{m}_{g_{k},t }(g_{k})$ in (\ref{eq.fourGaussApproximation}) by following (\ref{eq.y2ht}), (\ref{eq.hcom}), and (\ref{eq.fourGaussApproximation}). Then, $v_g$ is obtained by averaging the variances $v_{g_{k},t}$ in (\ref{eq.fourGaussApproximation}) over $k,t$, and the output MSE $v$ is finally given by (\ref{eq.SMD_Evo_V_gk}). The CSCE transfer function $v = f_{CSCE}(\tau)$ is thus obtained.

\subsection{Fixed Point of State Evolution and Convergence of JUICESD-RIGM }
The state evolution of the proposed JUICESD-RIGM algorithm is given by the recursion in (\ref{eq.OverallStateEvolution}). To explain the existence of the fixed point of (\ref{eq.OverallStateEvolution}) and the convergence of JUICESD-RIGM, we first show the monotonicity of the two transfer functions.
\begin{itemize}

\item For the SMD transfer function $\tau = f_{SMD}(v)$, we note that $v_g$ in (\ref{eq.SMD_Evo_V_gk}) is monotonically increasing with $v$, and that $v_g$ is used as the variance to generate the effective channel estimates $\{ \hat{g}_{k,t} \}$ in (\ref{eq.SMD_Evo_hat_gk}). Clearly, a smaller $v_g$ means more reliable $\{ \hat{g}_{k,t} \}$, and hence more reliable input messages to the SMD. Recall that the SMD follows the AMP principle. From the transfer function (\ref{eq.SSDEvolution1}), a more reliable \emph{a priori} distribution implies a more reliable output, i.e., a smaller output MSE $\tau$. In other words, the SMD function is monotonically increasing.

  \item For the CSCE transfer function $v = f_{CSCE}(\tau)$, a smaller $\tau$ implies more reliable input messages $\{ \hat{r}_{k,t} \}$ in (\ref{eq.CSCE_Evolution_model}), and so more reliable messages of the effective channels in (\ref{eq.hcom}), resulting in smaller $\{ v_{g_{k},t} \}$ in (\ref{eq.fourGaussApproximation}). Notice that $v_g$ is the average of $\{ v_{g_{k},t} \}$ and $v$ is monotonically increasing with $v_g$. Consequently, a smaller input MSE $\tau$ leads to a smaller output MSE $v$, and the CSCE transfer function is monotonically increasing.
\end{itemize}
Since both $f_{CSCE}(\tau)$ and $f_{SMD}(v)$ are monotonically increasing, the composite function $v = f_{CSCE}(f_{SMD}(v))$ is monotonically increasing.
We then show $v^{(q+1)} = f_{CSCE}(f_{SMD}(v^{(q)})), \, q \ge 0$ monotonically decreases as the iteration proceeds. As $f_{CSCE}(f_{SMD}(\cdot))$ is a monotonically increasing function, we only need to show $v^{(1)} \le v^{(0)}$, which can be verified by $v_g^{(1)} \le v_g^{(0)}$ since $v$ is monotonically increasing with $v_g$. In the first iteration, $v_g^{(0)}$ is initialized to be $\infty$ since there is no information fed back from SMD, which corresponds to $v^{(0)} = 1$ from (\ref{eq.SMD_Evo_V_gk}). After the first iteration, with certain information fed back from SMD, $v_{g_{k},t} < \infty$ in (\ref{eq.fourGaussApproximation}). Notice that $v_g$ is the average of $v_{g_{k},t}$ and so we have $v_g^{(1)} < \infty $, leading to $v^{(1)} < 1$ from (\ref{eq.SMD_Evo_V_gk}). Hence we have $v^{(1)} \le v^{(0)}$, and $v^{(q+1)} = f_{CSCE}(f_{SMD}(v^{(q)}))$ monotonically decreases as the iteration proceeds.
Finally, notice that the MSE $v$ is lower bounded by 0. From the monotone convergence theorem, $v$ always converges to a fixed point $v = f_{CSCE}(f_{SMD}(v))$; so does the proposed JUICESD-RIGM algorithm.

\begin{figure*}
\centering
\subfigure[Q-Qplot of the messages $\{\hat{g}_{k,t}\}$]{
\label{fig:gkt_qqplot_paper1}
\includegraphics[scale=0.5]{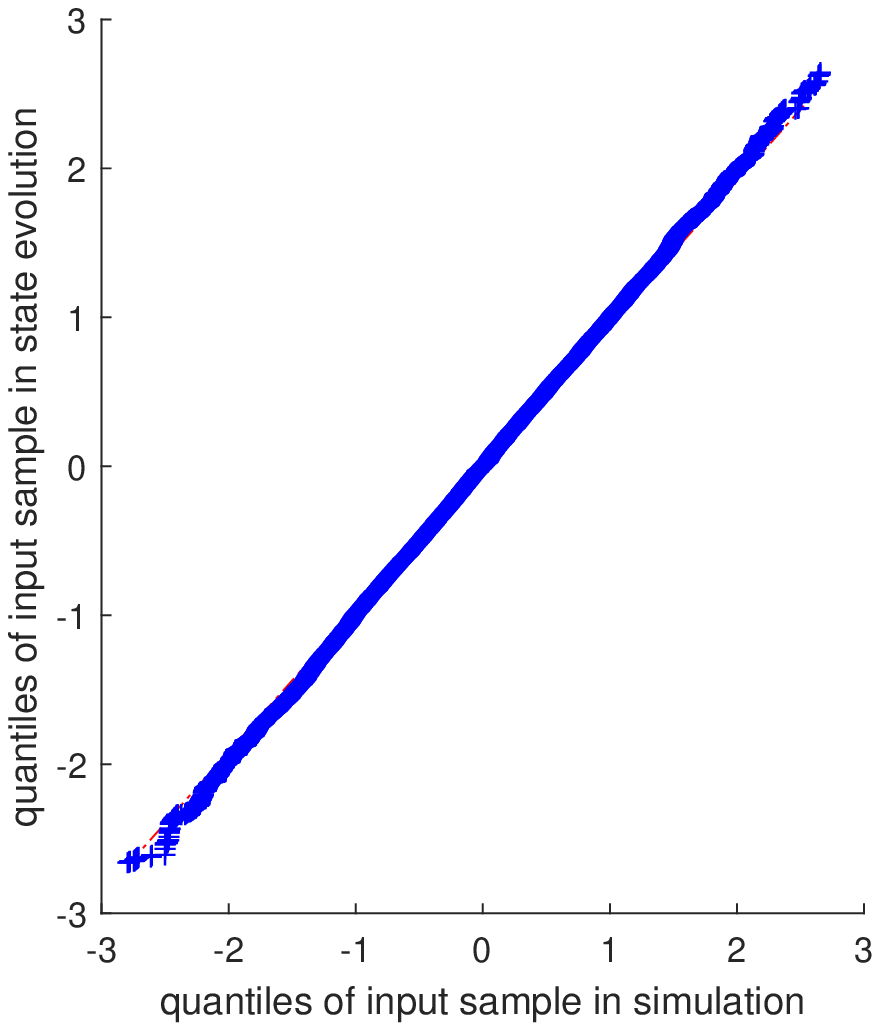}}
\vspace{0cm}
\subfigure[Q-Qplot of the messages $\{\hat{r}_{k,t}\}$]{
\label{fig:rkt_qqplot_paper1}
\includegraphics[scale=0.5]{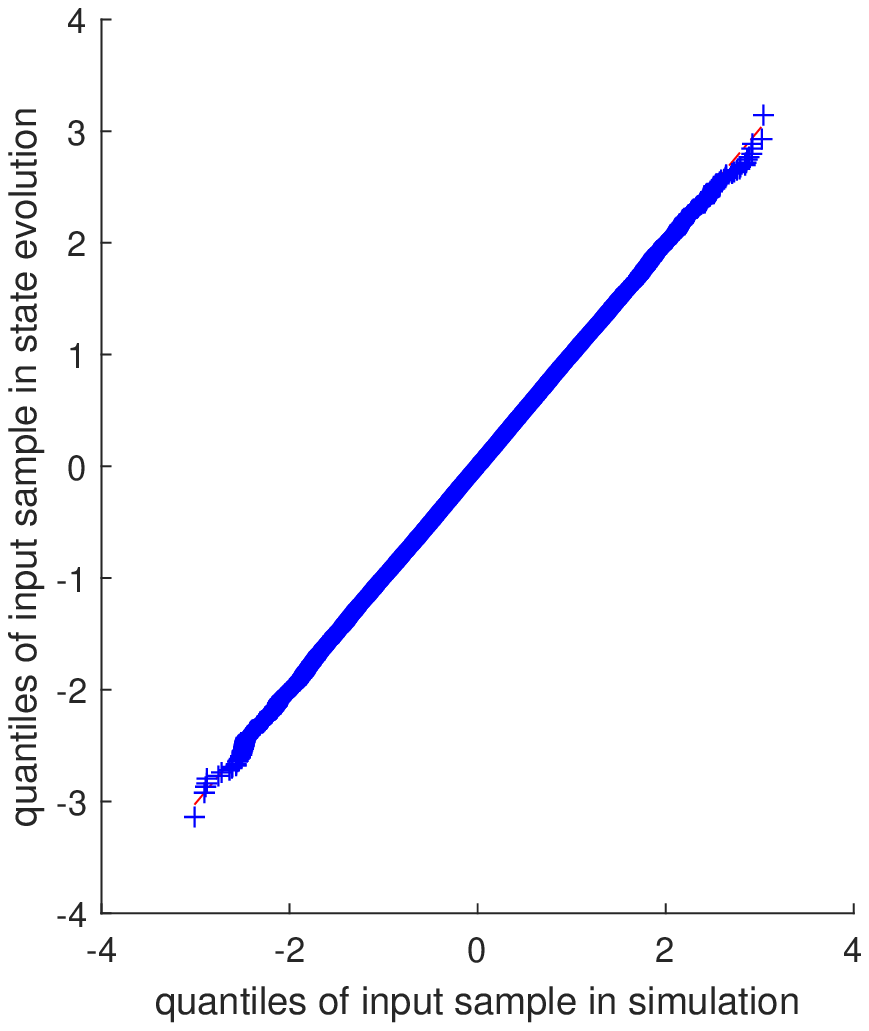}}
\label{Fig.QQplot K=2001}
\subfigure[Trajectory of JUICESD-RIGM.]{
\label{fig:StateEvolutionPa02}
\includegraphics[scale=0.5]{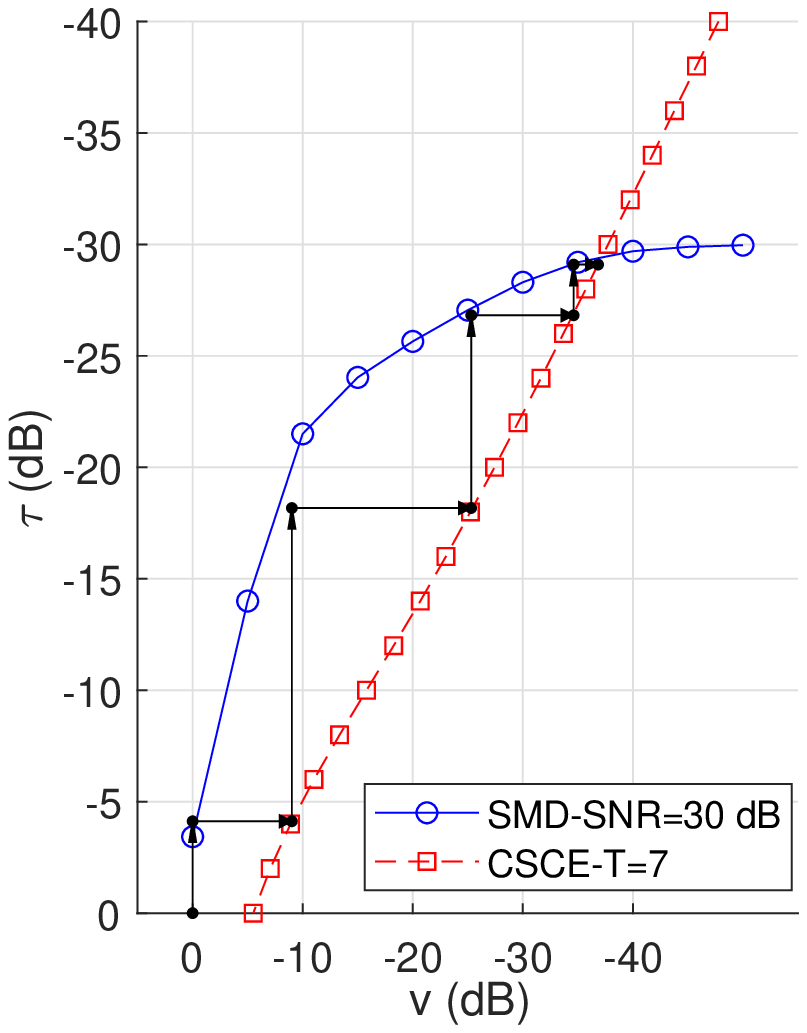}}
\caption{The accuracy of state evolution (SE).}
\end{figure*}

Fig. 6 provides some numerical results. Recall that the developed SE analysis is established based on the two assumptions (\ref{eq.SMD_Evo_hat_gk}) and (\ref{eq.CSCE_Evolution_model}) in modeling the input messages of CSCE and SMD respectively. In (\ref{eq.SMD_Evo_hat_gk}) and (\ref{eq.CSCE_Evolution_model}), we assume that (i) the estimation errors are Gaussian distributed; and (ii) the estimation errors are mutually independent and also independent of the variables to be estimated.
To verify assumption (i), we consider the Q-Q (quantile-quantile) plots of the input messages $\{\hat{g}_{k,t}\}$ to SMD and the Q-Q plot of the input messages $\{\hat{r}_{k,t}\}$ to CSCE in Fig. \ref{fig:gkt_qqplot_paper1} and Fig. \ref{fig:rkt_qqplot_paper1} respectively. We see that the Gaussian assumption (i) agrees well with the simulation results.
To further verify the independence assumption (ii), we compare the trajectory procedure obtained from simulations and that predicted by the state evolution in Fig. \ref{fig:StateEvolutionPa02}. From the figure, we see that the trajectory obtained by tracking the MSEs of the JUICESD-RIGM algorithm in simulations agrees well with the transfer functions of SE. This agreement is observed under all system parameters we simulated, which indirectly verifies the independence assumption (ii). It in turn verifies that the models in (\ref{eq.SMD_Evo_hat_gk}) and (\ref{eq.CSCE_Evolution_model}) are effective, so is the SE analysis.

The above SE can be used to predict the algorithm performance. To this end, both $f_{SMD}(\cdot)$ and $f_{CSCE}(\cdot)$ need to be pre-simulated and stored to form a look-up table. We will show by numerical results that the recursion in (\ref{eq.OverallStateEvolution}) accurately characterizes the performance of the JUICESD-RIGM algorithm.

\section{Numerical Results}\label{sec.Numerical}
In this section, we provide numerical results to verify the effectiveness of the proposed algorithms. The signal-to-noise ratio (SNR) is defined by $SNR=\frac{1}{N_0}$ (for normalized signal constellation ${\cal S}$ with $\sum \nolimits_{j=1}^{\vert {\cal S} \vert} \Vert s_j \Vert^2 / \vert {\cal S} \vert=1$). The data signals of all active users are QPSK modulated. In simulations, we mainly focus on the case that perfect power control is adopted to compensate for large scale fading such that $\beta_1 = \beta_2=\cdots=\beta_K \equiv 1$. We will briefly discuss the impact of large scale fading in the last subsection.

\subsection{Error Rate Performance}
We first define the activity error rate (AER) and the symbol error rate (SER) in the considered systems. For each frame, if the activities of all users are correct, the activity error is 0; otherwise, the activity error is 1. The AER is then obtained as the average of the activity errors over all simulated frames. The SER is calculated as follows. For inactive users, if its active state is judged correctly, all symbols are regarded as detected correctly; otherwise all incorrectly. For active users, we estimate their transmitted QPSK signals. A symbol is detected correctly only when both the user activity and the user symbol are judged correctly.

\begin{figure*}[!t]
\centering
\subfigure[AER versus SNR]{
\label{fig:AER-SNR for K=200}
\includegraphics[scale=0.5]{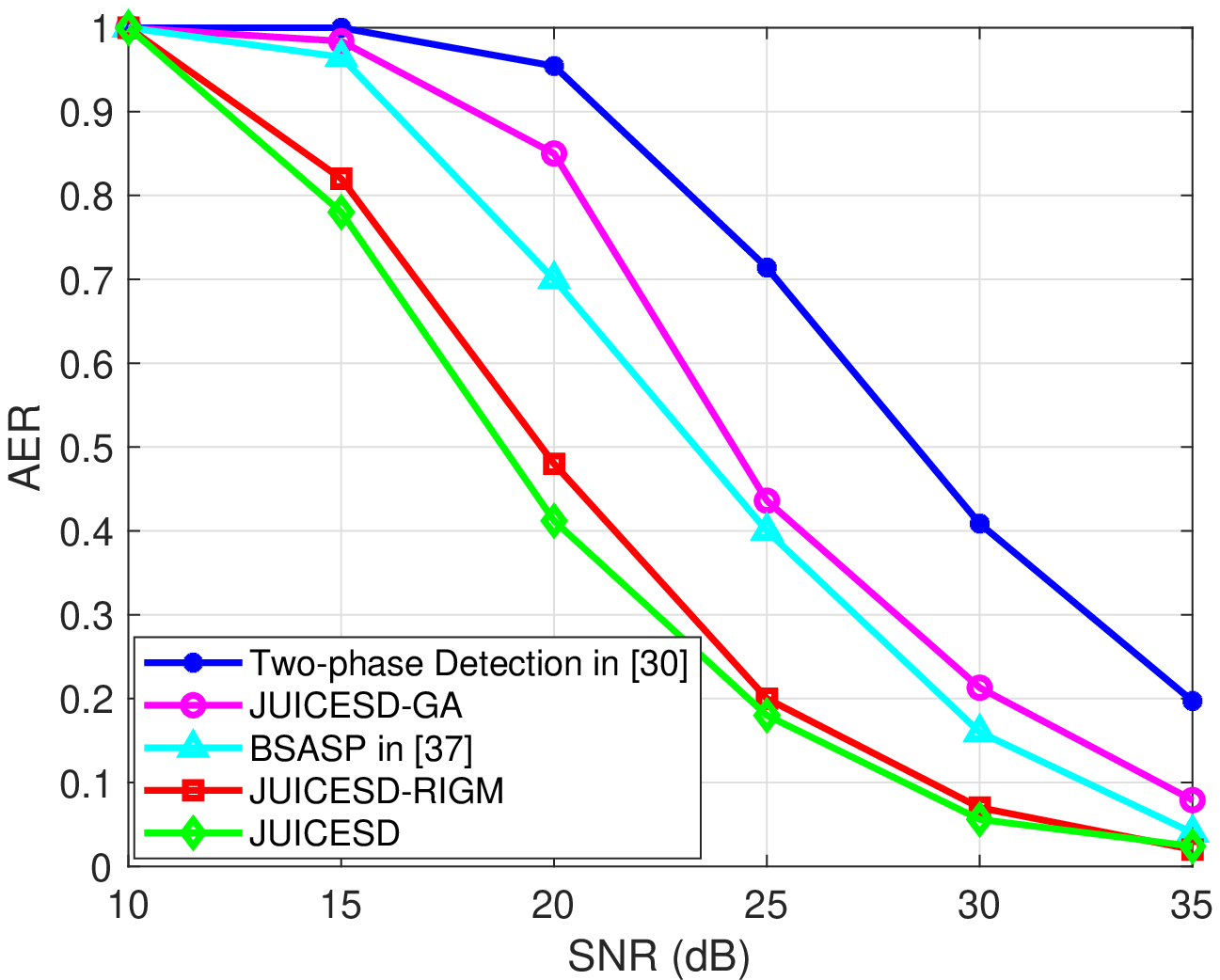}}
\vspace{0cm}
\subfigure[SER versus SNR]{
\label{fig:SER-SNR for K=200}
\includegraphics[scale=0.5]{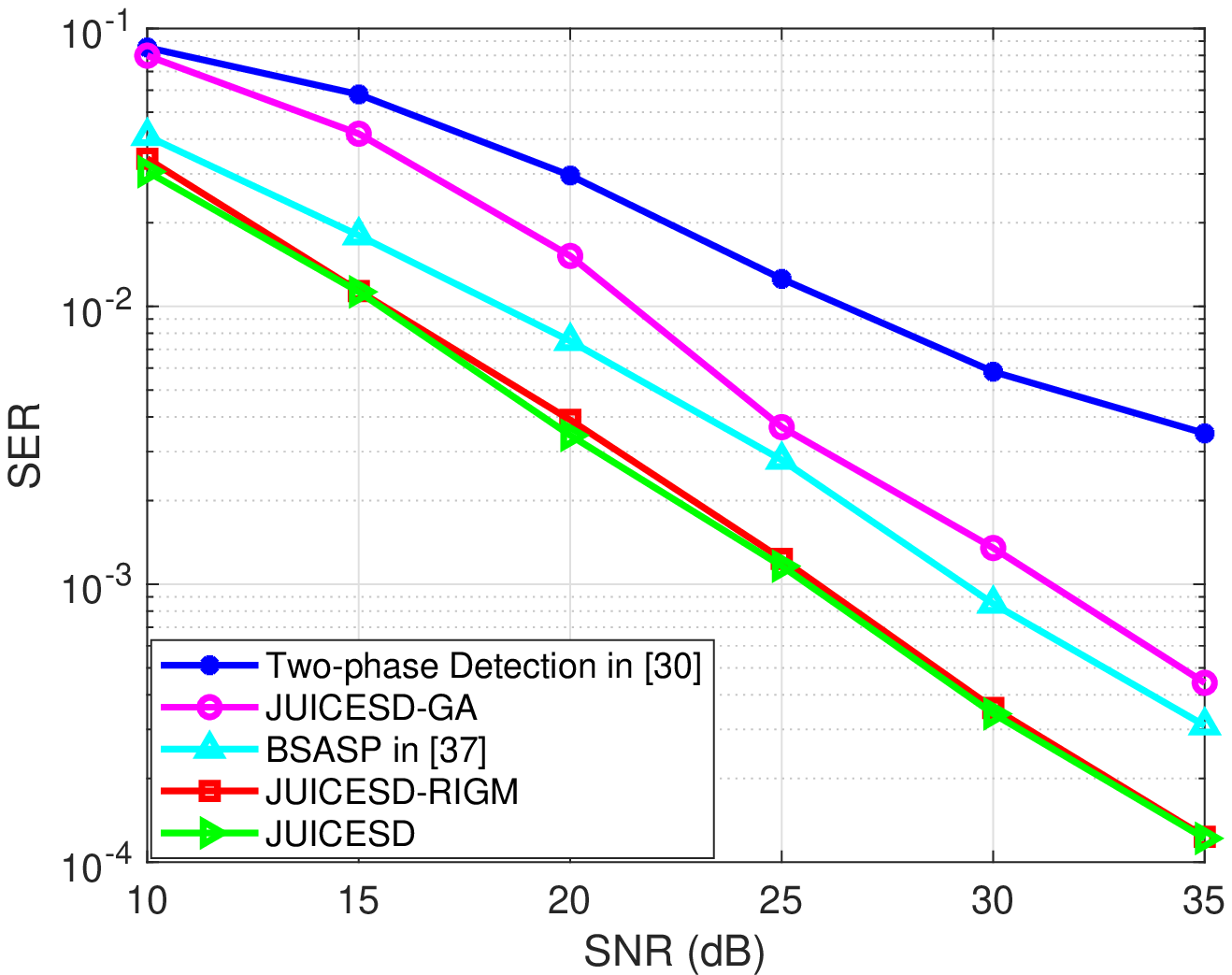}}
\caption{Performance versus SNR: $K=200$, $L=50$, $\lambda=0.1$, $T = 7$.}
\label{Fig.performance K=200}
\end{figure*}

\begin{figure*}
\centering
\subfigure[]{
\label{fig:SER-SNR for K=2000 pa=0.1}
\includegraphics[scale=0.54]{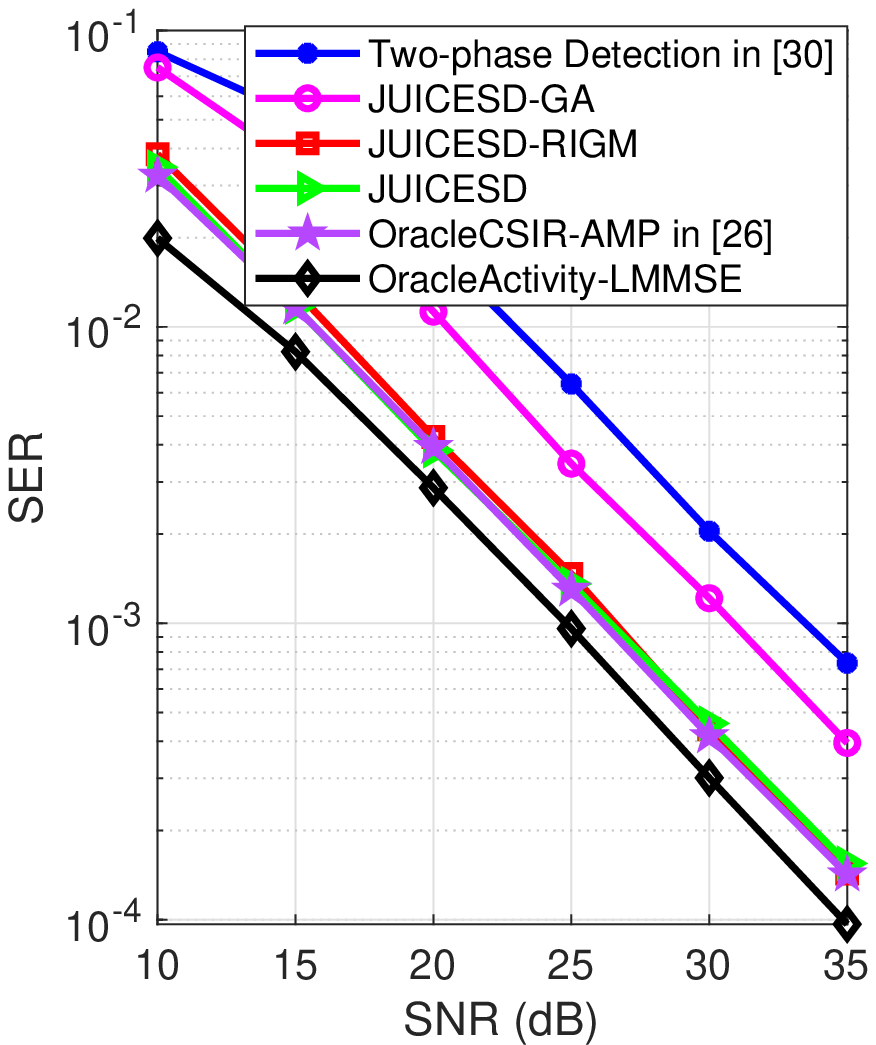}}
\vspace{0cm}
\subfigure[]{
\label{fig:SER-SNR for K=2000 pa=0.2}
\includegraphics[scale=0.54]{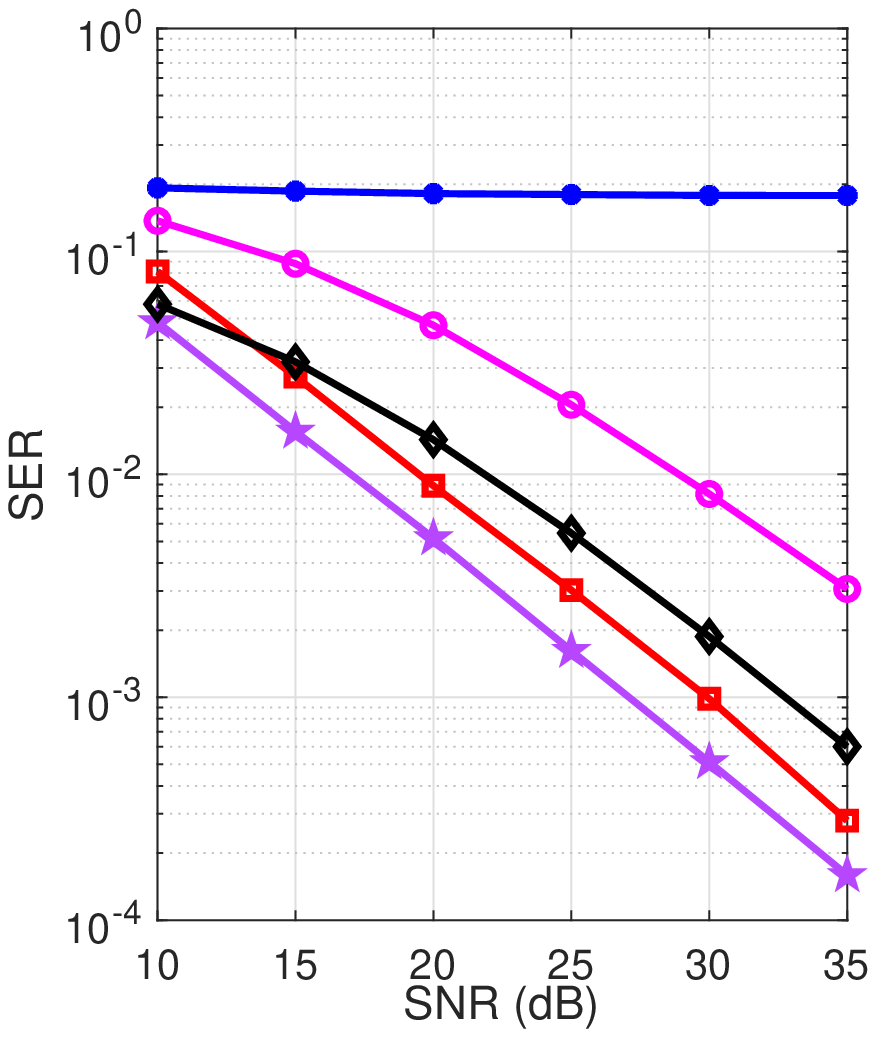}}
\subfigure[]{
\label{fig:SER-SNR for K=2000 pa=0.3}
\includegraphics[scale=0.54]{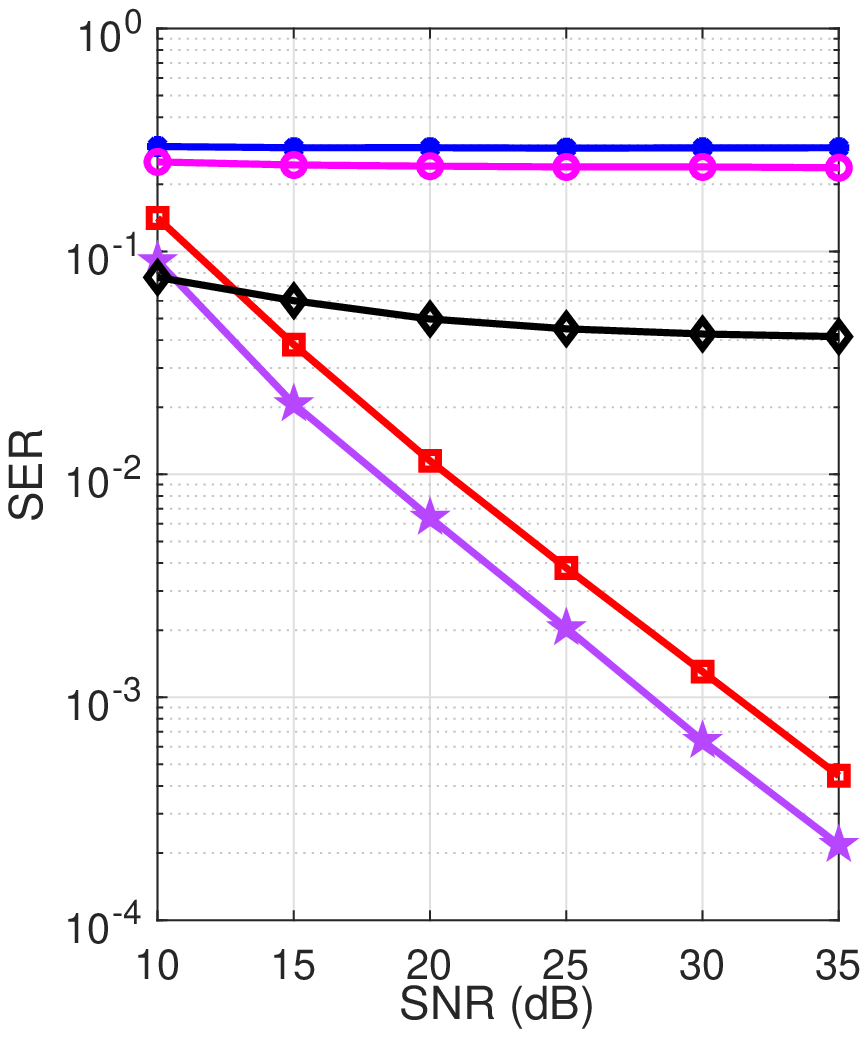}}
\caption{SER versus SNR: $K=2000$, $L=500$, $T = 7$. (a) $\lambda=0.1$, (b) $\lambda=0.2$, and (c) $\lambda=0.3$.}
\label{Fig.performance K=2000}
\end{figure*}

In Fig. \ref{Fig.performance K=200}, we compare the AER and SER performance of our proposed algorithms with the state-of-the-art BSASP algorithm in \cite{Du2018Joint}. For comparison, we also include the performance of the two-phase detection scheme \cite{Gabor2015Joint} in which the receiver first estimates the user activities and channel coefficients jointly, and then recovers the signals sent by the active users, both through AMP algorithms. In addition, we include the performance for JUICESD with message sampling in evaluating (\ref{eq.hcom}) (simply referred to as JUICESD) as a benchmark, and that for a so-called JUICESD-GA algorithm where the simple Gaussian approximation is used for message updates in the CSCE module of the proposed JUICESD algorithm.

From Fig. \ref{Fig.performance K=200}, we see that the two-phase detection scheme has a relatively high error floor while the BSASP and proposed JUICESD-RIGM algorithms work well. This demonstrates the benefit of joint user identification, channel estimation, and signal detection design. Furthermore, the proposed JUICESD-RIGM algorithm outperforms BSASP by about $3-4$ dB in terms of AER and SER. This is because BSASP suffers from non-orthogonal training, while the proposed JUICESD-RIGM algorithm relies on message passing principles to iteratively cancel/suppress the interference among users. Lastly, the JUICESD-RIGM algorithm and the (higher-complexity) JUICESD algorithm have similar performance. This implies that the proposed rotationally invariant Gaussian mixture model provides a good approximation for the complicated message involved in the joint design. On the other hand, the JUICESD-GA has an evident performance degradation due to its over-simplified Gaussian message structure.

We next apply the two-phase detection scheme, JUICESD-GA, and JUICESD-RIGM to a much larger system. We set the number of potential users $K=2000$, the length of spreading sequence $L=500$. We use the OracleActivity-LMMSE (in which user activity information is assumed to be perfectly known while both channel and data are estimated based on LMMSE principles) and the OracleCSIR-AMP in \cite{Wei2017Approximate} (in which perfect CSIR is assumed and the user activity and data are jointly detected by using AMP) as baselines. Note that we simulate the JUICESD only for the case of $\lambda=0.1$ due to its high complexity. Fig. \ref{Fig.performance K=2000} shows the SER performance of the considered algorithms. It can be seen that the proposed JUICESD-RIGM algorithm always outperforms the two-phase detection scheme, and it achieves a substantial performance improvement over JUICESD-GA. In addition, JUICESD-RIGM only has a small SER gap to OracleCSIR-AMP especially when the user activity probability $\lambda$ is small. This implies that JUICESD-RIGM can estimate the channel accurately through joint detection. Note that JUICESD-RIGM outperforms OracleActivity-LMMSE when $\lambda$ is large (i.e., $\lambda=0.2, 0.3$). This is possible since LMMSE is a linear detection algorithm that strictly sub-optimal when a finite constellation alphabet is employed in modulation.

\begin{figure}
\centering
\subfigure[AER performance]{
\label{fig:AER_pilotNUMPaper}
\includegraphics[scale=0.5]{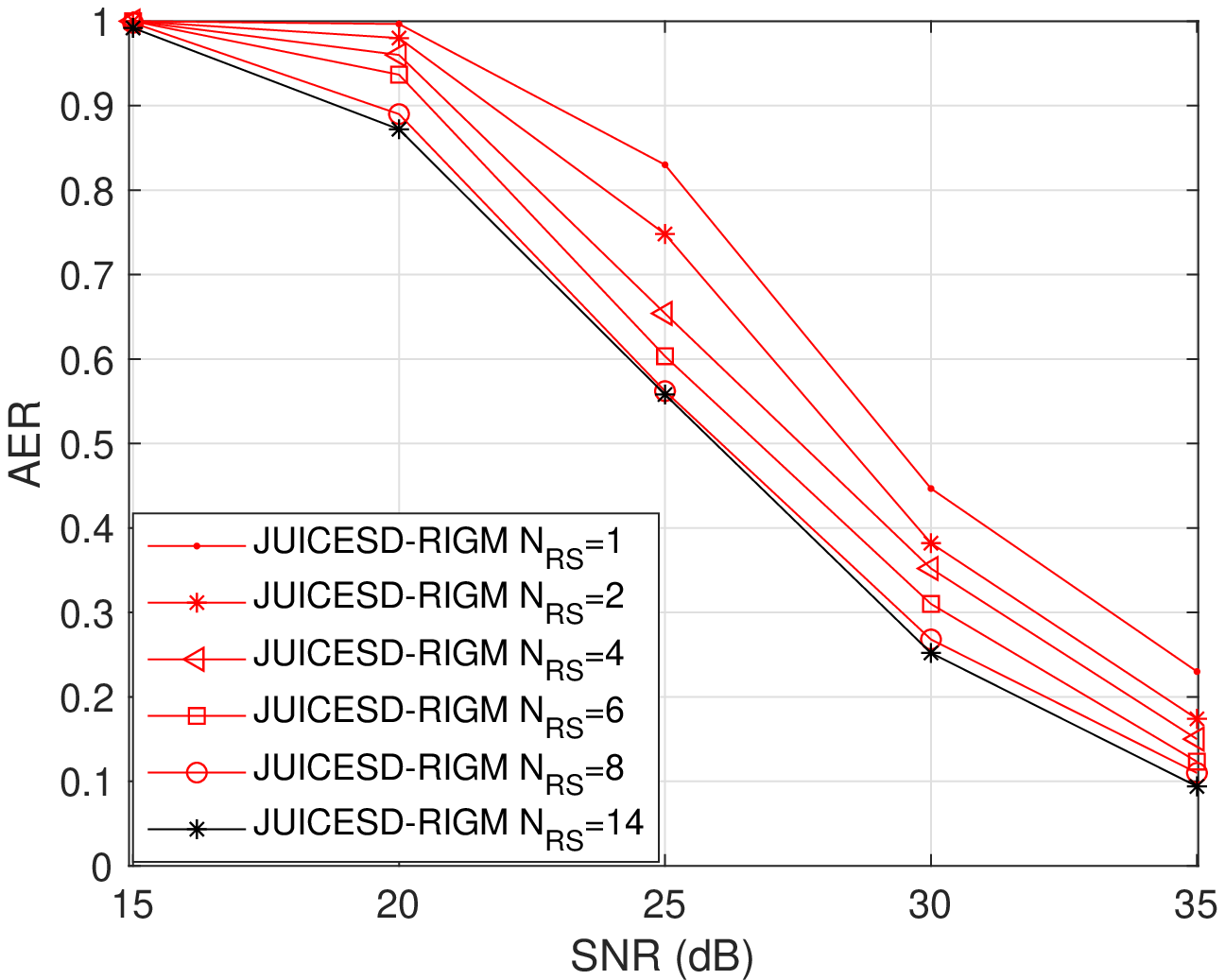}}
\vspace{0cm}
\subfigure[SER performance]{
\label{fig:SER_pilotNUMPaper}
\includegraphics[scale=0.5]{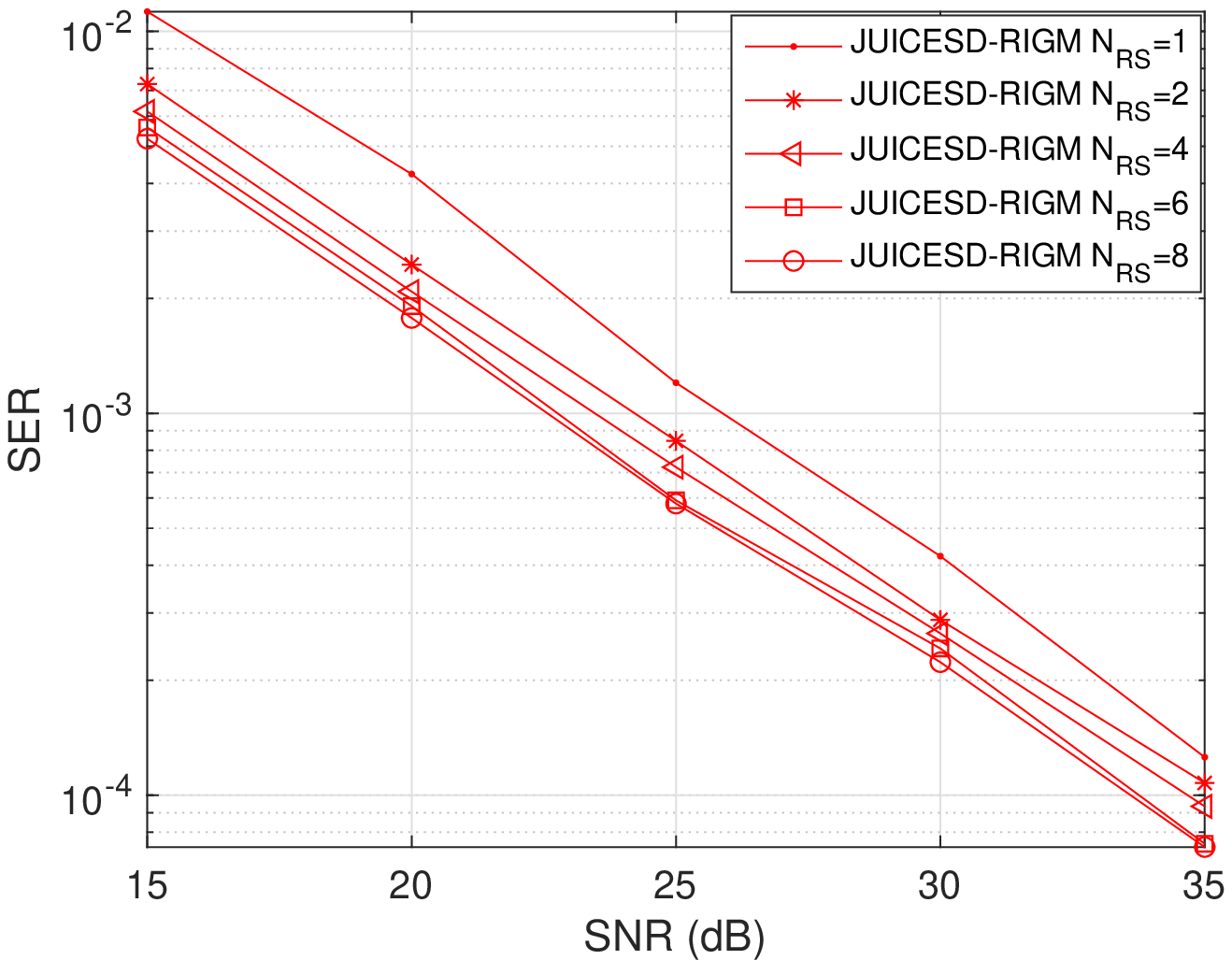}}
\caption{Performance versus SNR with different number of reference signals $N_{RS}$. $T = 14$. }
\label{Fig.PerformancePilotNum}
\end{figure}

Fig. \ref{fig:AER_pilotNUMPaper} and Fig. \ref{fig:SER_pilotNUMPaper} depict the effect of increasing the number of reference signals $N_{RS}$ on AER and SER. It can be seen that as $N_{RS}$ increases, the system performance improves gradually (at the cost of a larger system overhead). Such a performance improvement becomes marginal when $N_{RS} \ge 8$. That is, further increasing the number of reference signals $N_{RS}$ has a negligible effect on the system performance.

\subsection{Phase Transition Performance}
\begin{figure}[!t]
\begin{center}
\includegraphics[scale=0.5]{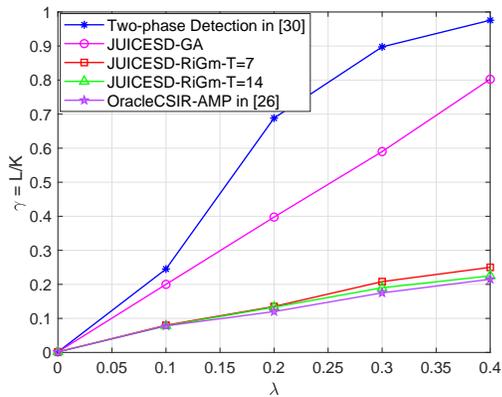}
\caption{Phase transition performance with SER threshold $SER_{th} = 10^{-3}$. SNR = 35 dB.}
\label{fig:PhaseTrans}
\end{center}
\end{figure}

Fig. \ref{fig:PhaseTrans} shows the phase transition performance of the proposed JUICESD-RIGM algorithm, where $\lambda$ is the user activity probability and $\gamma = L/K$. A transmission is declared as a success if its SER is lower than a given threshold $SER_{th}$; otherwise the transmission is declared as a failure. It is observed from Fig. \ref{fig:PhaseTrans} that the system load supported by JUICESD-RIGM is much larger than the other two baseline algorithms. For example, at $\gamma = 0.1$, JUICESD-RIGM can support activity probability of $\lambda = 0.13$, which is much greater than $\lambda = 0.05$ for JUICESD-GA and $\lambda = 0.03$ for two-phase detection.

It can be seen in the phase transition that JUICESD-RIGM can achieve a good performance even when $\lambda > \gamma$, and closely approaches the OracleCSIR-AMP curve throughout the considered range of $\lambda$. Note that $\lambda > \gamma$ implies the number of active users supported by the system exceeds the spreading length $L$. This is possible because the user activities are fixed across $T$ slots (effectively increasing the number of observations by a factor of $T=7,14$) and because of non-ideal coding (i.e., QPSK with $SER_{th}=10^{-3}$).

From Fig. \ref{fig:PhaseTrans}, the gap between our algorithm and OracleCSIR-AMP increases slightly as $\lambda$ increases. The reason is that as $\lambda$ increases, the interference between users increases, which makes the channel estimation difficult. In addition, the gap between our algorithm and OracleCSIR-AMP reduces as the frame length $T$ increases. The reason is that, as $T$ increases, more partially decoded data symbols can be used as pilot to enhance channel estimation in the iterative channel estimation and signal detection process, thereby yielding a better system performance.

\subsection{State Evolution}
\begin{figure}[!t]
\centering
\subfigure[MSE Performance]{
\label{fig:StateEvolutionMSE}
\includegraphics[scale=0.5]{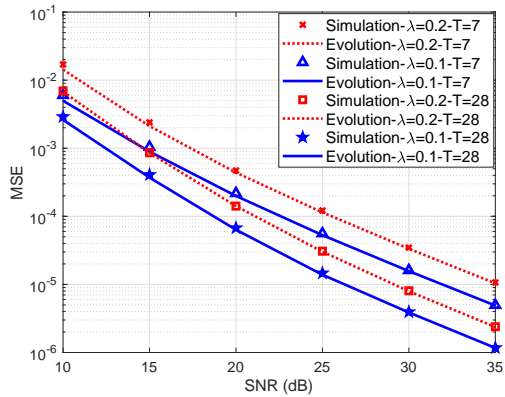}}
\vspace{0cm}
\subfigure[SER Performance]{
\label{fig:StateEvolutionSER}
\includegraphics[scale=0.5]{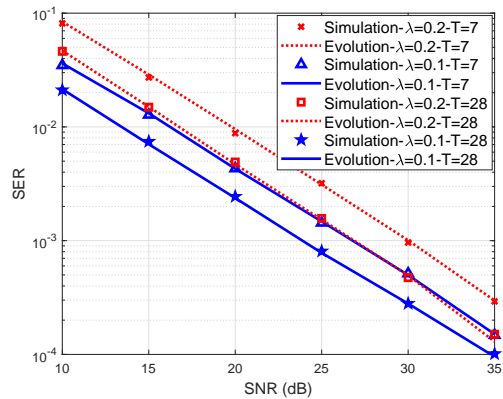}}
\caption{Performance comparison between state evolution and simulation; $K = 2000$ and $L = 500$.}
\label{Fig.StateEvolutionperformance}
\end{figure}

We evaluate the accuracy of the state evolution of JUICESD-RIGM in this subsection.
Figs. \ref{fig:StateEvolutionMSE} and \ref{fig:StateEvolutionSER} show the accuracy of the state evolution in predicting the MSE and SER performance. By defining $\bm{g}=[g_1,g_2, \cdots, g_K]^T$, we calculate $MSE=\frac{1}{K} \Vert \bm{g} - \hat{\bm{g}}\Vert_2^2 $, where $\hat{\bm{g}}$ is the estimate of $\bm{g}$ given by the JUICESD-RIGM algorithm. We compare the simulated MSE and SER performance with the prediction by the state evolution. From Figs. \ref{fig:StateEvolutionMSE} and \ref{fig:StateEvolutionSER}, the performance predicted by the state evolution is very close to that by simulation. Hence, the state evolution can track the performance of JUICESD-RIGM accurately.

\vspace{-0.3cm}
\subsection{Large Scale Fading}
\begin{figure}[!t]
\centering
\subfigure[]{
\label{fig:LargeFading01}
\includegraphics[scale=0.5]{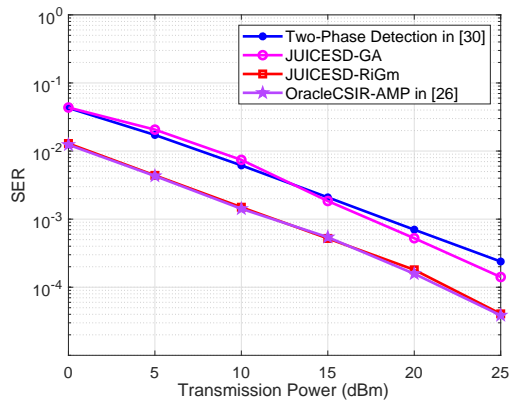}}
\vspace{0.3cm}
\subfigure[]{
\label{fig:LargeFading02}
\includegraphics[scale=0.5]{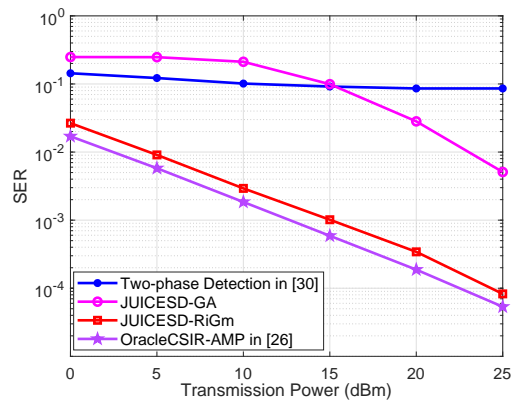}}
\caption{Performance with large scale fading: $K=2000$, $L=500$, $T=7$. (a) $\lambda=0.1$ and (b) $\lambda=0.2$.}
\label{Fig.LargeScaleFading}
\end{figure}

In this subsection, we discuss the impact of large scale fading on the system performance. We assume that $\{\beta_k\}$ are known at the receiver. This implies that the channels of different users are generally non-i.i.d. Let $d_k$ denote the distance between user $k$ and the AP. We assume that $\{ d_k \}$ are uniformly distributed in the range of $[0.05km, 1km]$, i.e., $d_k \sim U(0.05,1)$ where $U(a,b)$ denotes the uniform distribution with the minimum value $a$ and the maximum value $b$. The path loss model of the wireless channel for user $k$ is given as $\beta_k= - 128.1 - 36.7 {\rm log}_{10}(d_k)$ in dB. The bandwidth of the wireless channel is $1$MHz. The power spectral density of the AWGN at the AP is $-169$dBm/Hz.
Fig. \ref{Fig.LargeScaleFading} shows the SER performance with different user activity probabilities. Similar observations as in Section VI-A can be made. This verifies the effectiveness of the proposed JUICESD-RIGM algorithm in a more practical scenario.

\vspace{-0.3cm}

\subsection{Complexity}
We next compare the complexity of the proposed algorithm with existing alternatives. Table \ref{Tab:Complexity} provides the expressions of their complexity order and the number of real multiplications required by the algorithms in one iteration, where $\rho$ is the sparse level in BSASP \cite{Du2018Joint}.

It can be observed that the complexities of JUICESD-RIGM and two-phase detection \cite{Gabor2015Joint} are comparable while that of BSASP is much higher. Recall that the two-phase detection algorithm estimates the user activities, channel coefficients and signals in two separated phases, implying that its low complexity is obtained at the expense of a significant performance loss, as demonstrated in Figs. \ref{Fig.performance K=200}, \ref{Fig.performance K=2000}, \ref{fig:PhaseTrans}, \ref{Fig.LargeScaleFading}.
\begin{table*}
\caption{Complexity Comparison}
\footnotesize
\renewcommand{\arraystretch}{1.3}
\label{Tab:Complexity}
\centering
\begin{tabular}{c||c||c}
\hline
\bfseries Algorithms & \bfseries Complexity order & \bfseries Number of real multiplications \\
\hline
JUICESD-RIGM &  ${\cal O}(KTL) + {\cal O}(KT^2 \vert {\cal S} \vert)$ & $\left[ (22T-3) \vert {\cal S} \vert + 12 \vert \Omega_{\cal S} \vert + 2\frac{\vert {\cal S} \vert}{\vert \Omega_{\cal S} \vert} + 2T +27 \right] KT +6LKT + 10LT$\\
\hline
BSASP & ${\cal O}(KLT^2) + {\cal O}(LT^3 \rho^2) + {\cal O}(T^3 \rho^3)$ & $2T^3 \rho^3 + 4LT^3\rho^2 + 6LKT^2 +6KT + 10K$  \\
\hline
Two-phase Detection & ${\cal O}(KT \vert {\cal S} \vert) + {\cal O}(KTL)$ & $(21 \vert {\cal S} \vert + 22)KT +6LKT + 10LT$\\
\hline
\end{tabular}
\end{table*}
\section{Conclusions and Future Works}\label{sec.Conclusion}
We proposed a novel JUICESD framework for mMTC applications. A low-complexity yet efficient algorithm named JUICESD-RIGM was developed based on judicious iterative receiver design, sophisticated message passing principles, and accurate rotationally invariant Gaussian mixture message structure. Furthermore, we established the state evolution analysis to predict the performance of JUICESD-RIGM. Numerical results demonstrate that JUICESD-RGMA achieves a significant performance gain over the state-of-the-art algorithms and even outperforms LMMSE receivers with oracle user activity information. The complexity of JUICESD-RIGM is also low; hence it is especially suitable for machine type communications with a massive number of potential devices and short packets. In addition, the performance of JUICESD-RIGM predicted by the state evolution is very close to that by the simulation, which provides insights for future system design and optimization.

This paper is focused only on the single-antenna configuration at the receiver, which is suitable for mMTC applications with short packet length and low transmission rate. Multi-antenna configuration will be an interesting extension to support more users and/or higher transmission rates. To exploit the potential structural information of the multi-antenna channels, such as sparsity in the angular domain, more advanced designs can be involved. This is, however, beyond the scope of this paper and will be pursued in our future work.

\section*{Appendix A: Proof of Lemma 1}
%\begin{IEEEproof}
We first prove (\ref{eq.reducedmean}). To this end, it suffices to show $\hat{g}_{k,i'}^{(T'+1)} = \hat{g}_{k,i}^{(T'+1)} e^{\text{j} (i'-i) \theta_0}$ for any $i'=1,\cdots,\vert \Omega_{S} \vert$. We note from (\ref{combined_mean_variance_multi_Gaussian}) that if for any $i'=1,\cdots,\vert \Omega_{\cal S} \vert$ and $j'$ satisfying $ s_{j'} = s_{j} e^{\text{j} (i'-i) \theta_0}$, the following equalities hold:
\begin{equation} \label{eq.conditions}
w_{i',j'}^{(T')} = w_{i,j}^{(T')} \  \text{and} \ \mu_{i',j'}^{(T')} = \mu_{i,j}^{(T')} e^{\text{j} (i'-i) \theta_0},
\end{equation}
then
\begin{equation}
\begin{split}
\hat{g}_{k,i'}^{(T'+1)} & = \sum_{j'=1}^{\vert {\cal S} \vert} w_{i',j'}^{(T')}\mu_{i',j'}^{(T')} \\ & = \sum_{j=1}^{\vert {\cal S} \vert} w_{i,j}^{(T')}\mu_{i,j}^{(T')} e^{\text{j} (i'-i) \theta_0} \\
&= \hat{g}_{k,i}^{(T'+1)} e^{\text{j} (i'-i) \theta_0}.
\end{split}
\end{equation}
What remains is to verify (\ref{eq.conditions}). Recall from (\ref{eq.fourGaussApproximation_temp2}) that $\hat{g}_{k,i'}^{(T')} = \hat{g}_{k,i}^{(T')} e^{\text{j} (i'-i) \theta_0}$. From (\ref{weights_multi_Gaussian}), we have

\begin{equation}\label{weights_multi_Gaussian_ij2}
\begin{split}
w_{i',j'}^{(T')} & \propto \frac{{\rm exp}\left( - \frac{\Vert \hat{g}_{k,i}^{(T')} e^{\text{j} (i'-i) \theta_0} -\hat{r}_{k,T^\prime+1}  /  s_{j'} \Vert^2 }{ v_{g_{k}}^{(T')} + v_{r_{k,T^\prime+1}}  / \Vert s_{j'} \Vert^2}  \right)}{\pi ( v_{g_{k}}^{(T')} + v_{r_{k,T^\prime+1}}  / \Vert s_{j'} \Vert^2)} \\
& = \frac{{\rm exp}\left( - \frac{\Vert \hat{g}_{k,i}^{(T')} -\hat{r}_{k,T^\prime+1}  / \left( s_{j'} e^{\text{j} (i-i') \theta_0} \right) \Vert^2 }{ v_{g_{k}}^{(T')} + v_{r_{k,T^\prime+1}}  / \Vert s_{j'} e^{\text{j} (i-i') \theta_0} \Vert^2}  \right)}{\pi ( v_{g_{k}}^{(T')} + v_{r_{k,T^\prime+1}}  / \Vert s_{j'} e^{\text{j} (i-i') \theta_0}\Vert^2)}  \\
&= w_{i,j}^{(T')}.
\end{split}
\end{equation}

From (\ref{means_variances_multi_Gaussian}), we have

\begin{equation}\label{means_variances_multi_Gaussian_proof}
\begin{split}
\mu_{i',j'}^{(T')} &= \frac{ \frac{\hat{g}_{k,i}^{(T')} e^{\text{j} (i'-i) \theta_0} v_{r_{k,T^\prime+1}}}{\Vert s_{j'} \Vert^2}   + \frac{v_{g_{k}}^{(T')} \hat{r}_{k,T^\prime+1}}{s_{j'}}  }{ v_{g_{k}}^{(T')} + v_{r_{k,T^\prime+1}}  / \Vert s_{j'} \Vert^2 } \\
 & = \frac{ \frac{\hat{g}_{k,i}^{(T')}  v_{r_{k,T^\prime+1}}}{\Vert s_{j'} e^{\text{j} (i-i') \theta_0} \Vert^2}  + \frac{v_{g_{k}}^{(T')} \hat{r}_{k,T^\prime+1} }{ s_{j'} e^{\text{j} (i-i') \theta_0} }  }{ v_{g_{k}}^{(T')} + v_{r_{k,T^\prime+1}}  / \Vert s_{j'} e^{\text{j} (i-i') \theta_0} \Vert^2 } e^{\text{j} (i'-i) \theta_0} \\
 & = \mu_{i,j}^{(T')} e^{\text{j} (i'-i) \theta_0}.
\end{split}
\end{equation}
Similarly, we can prove (\ref{eq.reducedvar}). The lemma then readily follows.

\bibliographystyle{IEEEtran}
\bibliography{IEEEabrv,reference}

\begin{IEEEbiography}[{\includegraphics[width=1in,height=1.25in,clip,keepaspectratio]{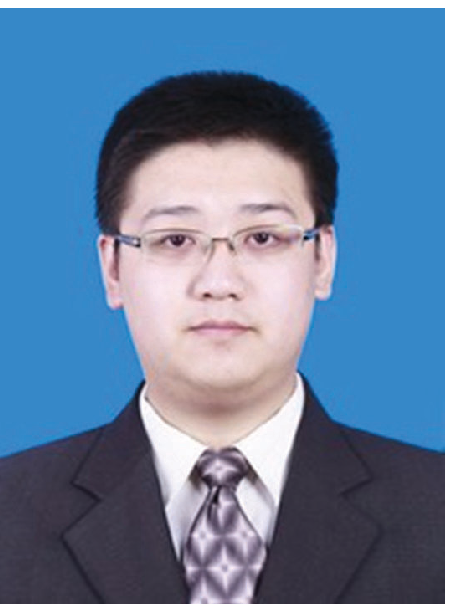}}]{Shuchao Jiang} is currently pursing the Ph.D. degree with the Department of Communication Science and Engineering, Fudan University, China. His research interests include wireless communication and signal processing.
\end{IEEEbiography}

\begin{IEEEbiography}[{\includegraphics[width=1in,height=1.25in,clip,keepaspectratio]{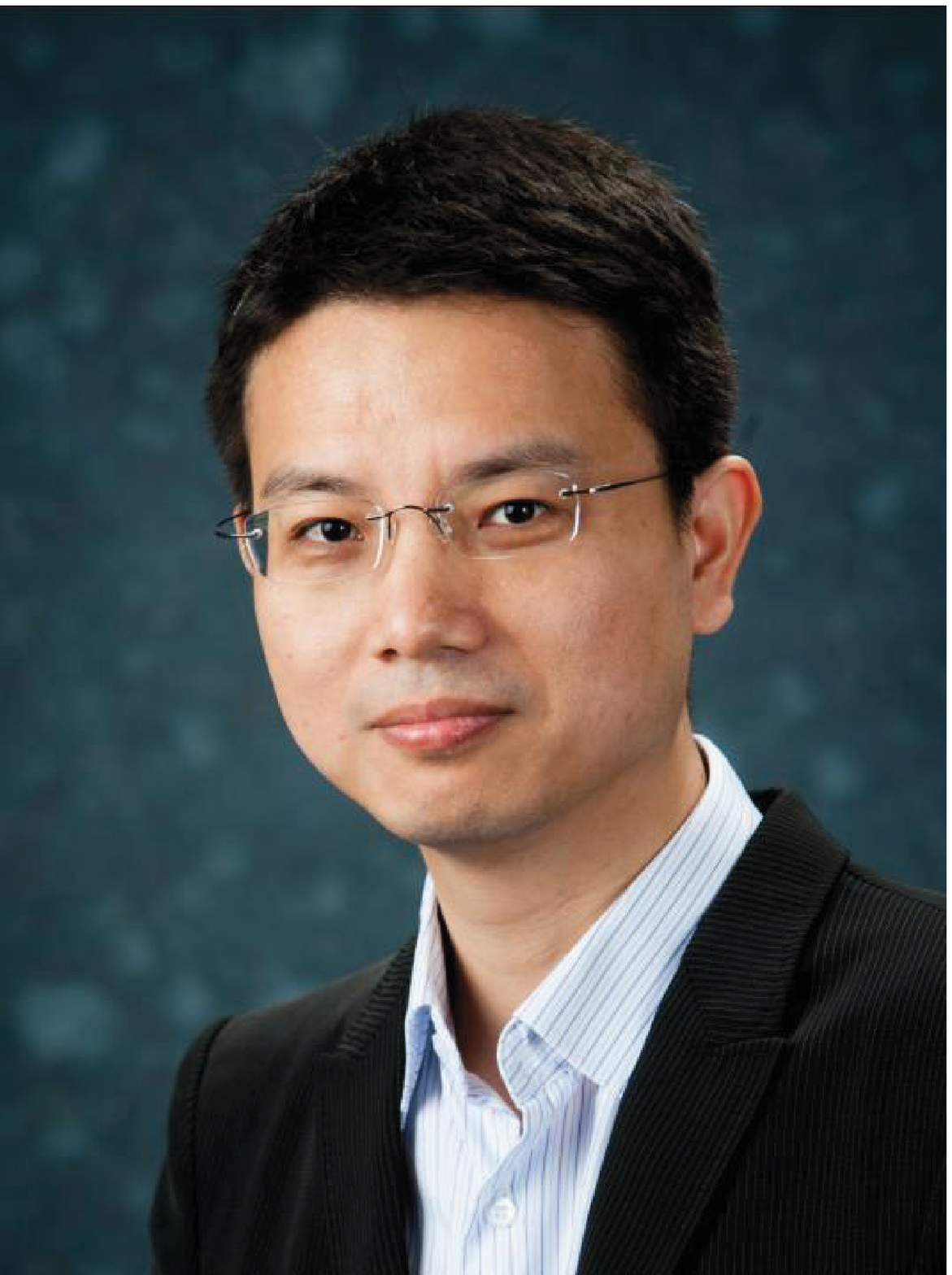}}]{Xiaojun Yuan}(S'04-M'09-SM'15) received the Ph.D. degree in Electrical Engineering from the City University of Hong Kong in 2008. From 2009 to 2011, he was a research fellow at the Department of Electronic Engineering, the City University of Hong Kong. He was also a visiting scholar at the Department of Electrical Engineering, the University of Hawaii at Manoa in spring and summer 2009, as well as in the same period of 2010. From 2011 to 2014, he was a research assistant professor with the Institute of Network Coding, The Chinese University of Hong Kong. From 2014 to 2017, he was an assistant professor with the School of Information Science and Technology, ShanghaiTech University. He is now a professor with the Center for Intelligent Networking and Communications, the University of Electronic Science and Technology of China.

His research interests cover a broad range of signal processing, machine learning, and wireless communications, including but not limited to multi-antenna and cooperative communications, sparse and structured signal recovery, Bayesian approximate inference, network coding, etc. He has published over 160 peer-reviewed research papers in the leading international journals and conferences in the related areas. He has served on a number of technical programs for international conferences. He is an editor of the IEEE Transactions on Communications since 2017, and also an editor of the IEEE Transactions on Wireless Communications since 2018. He was a co-recipient of the Best Paper Award of IEEE International Conference on Communications (ICC) 2014, and also a co-recipient of the Best Journal Paper Award of IEEE Technical Committee on Green Communications and Computing (TCGCC) 2017.
\end{IEEEbiography}

\begin{IEEEbiography}[{\includegraphics[width=1in,height=1.25in,clip,keepaspectratio]{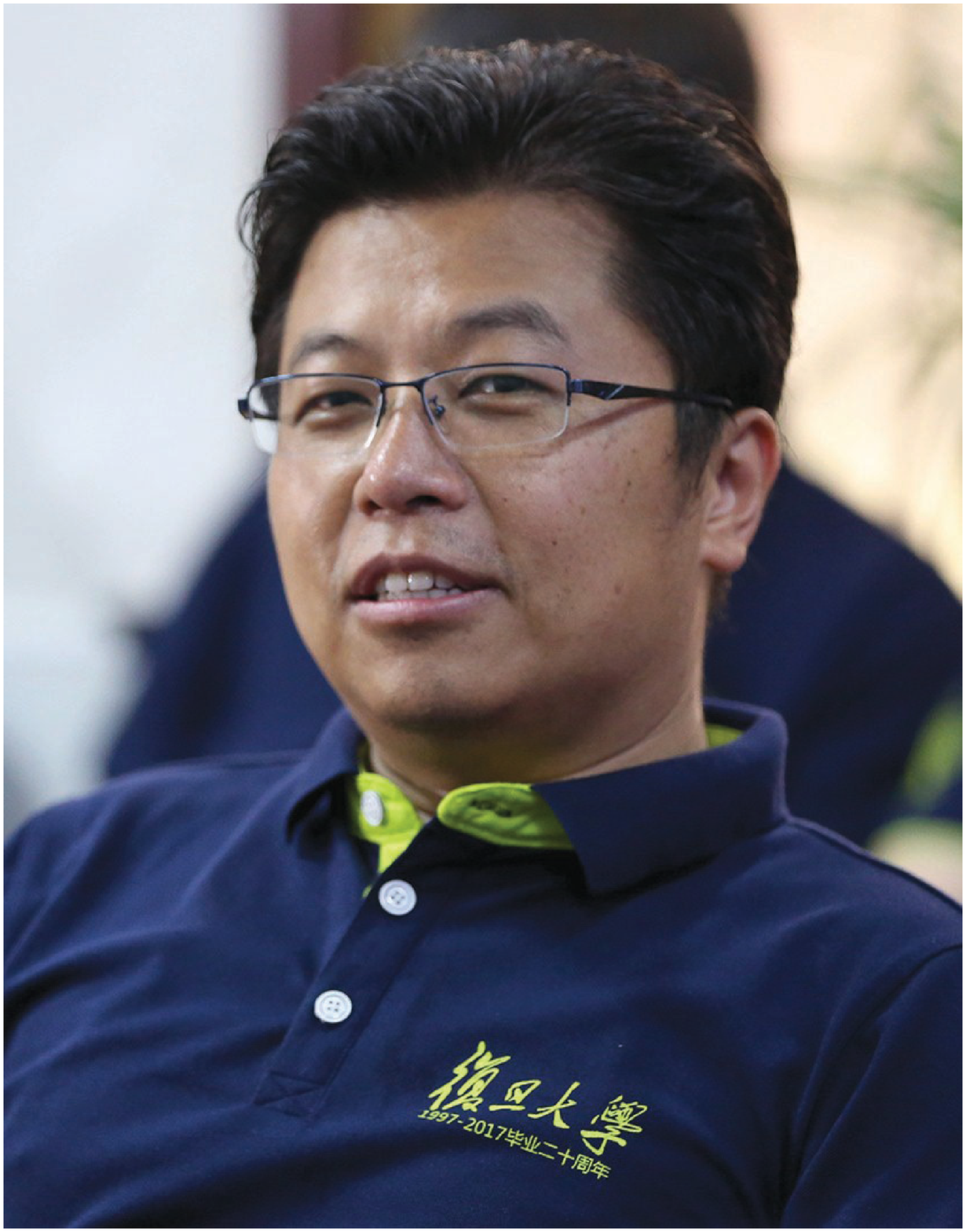}}]{Xin Wang}(SM'09) received the B.Sc. and M.Sc.degrees from Fudan University, Shanghai, China, in 1997 and 2000, respectively, and the Ph.D. degree from Auburn University, Auburn, AL, USA, in 2004, all in electrical engineering. From September 2004 to August 2006, he was a Postdoctoral Research Associate with the Department of Electrical and Computer Engineering, University of Minnesota, Minneapolis. In August 2006, he joined the Department of Electrical Engineering, Florida Atlantic University, Boca Raton, FL, USA, as an Assistant Professor, then was promoted to a tenured Associate Professor in 2010. He is currently a Distinguished Professor and the Chair of the Department of Communication Science and Engineering, Fudan University, China. His research interests include stochastic network optimization, energy-efficient communications, cross-layer design, and signal processing for communications. He served as an Associate Editor for the IEEE Transactions on Signal Processing, as an Editor for the IEEE Transactions on Vehicular Technology, and as an Associate Editor for the IEEE Signal Processing Letters. He currently serves as a Senior Area Editor for the IEEE Transactions on Signal Processing and as an Editor for the IEEE Transactions on Wireless Communications. He is an IEEE Distinguished Lecturer for the Vehicular Technology Society.
\end{IEEEbiography}

\begin{IEEEbiography}[{\includegraphics[width=1in,height=1.25in,clip,keepaspectratio]{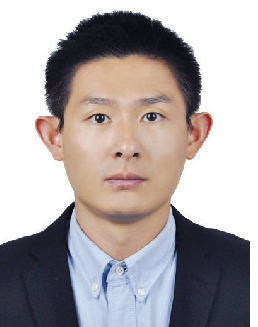}}]{Chongbin Xu}(M'16) received his B.S. degree in Information Engineering from Xi¡¯an Jiaotong University in 2005 and PhD degree in Information and Communication Engineering from Tsinghua University in 2012. From January 2012 to December 2014, he was a research fellow at the Department of Electronic Engineering, the City University of Hong Kong. Since December 2014, he has been with the Department of Communication Science and Engineering, Fudan University, China. His research interests are in the areas of signal processing and communication theory, including linear precoding, iterative detection, and multiple access techniques.
\end{IEEEbiography}

\begin{IEEEbiography}[{\includegraphics[width=1in,height=1.25in,clip,keepaspectratio]{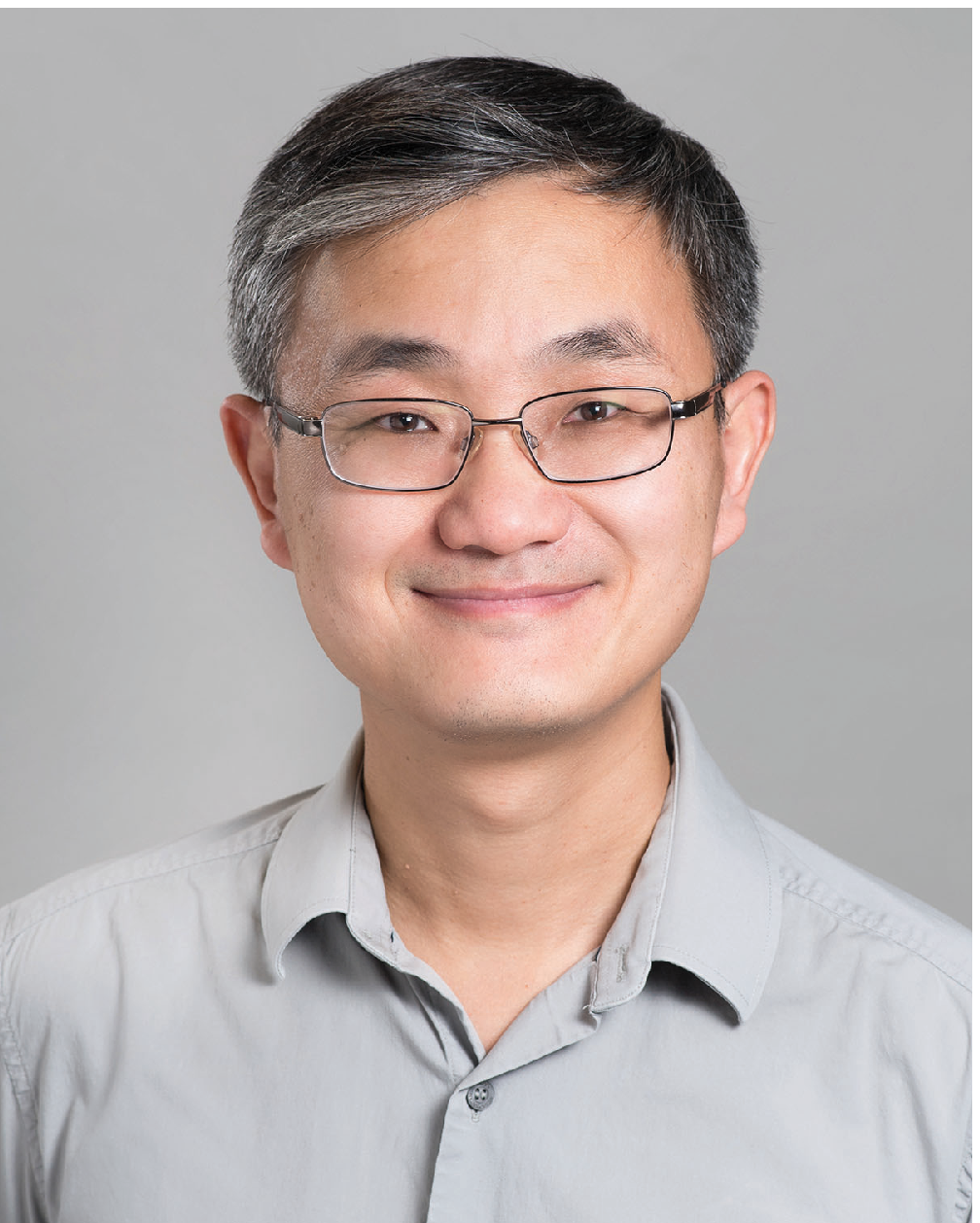}}]{Wei Yu}(S'97-M'02-SM'08-F'14) received the B.A.Sc. degree in Computer Engineering and Mathematics from the University of Waterloo, Waterloo, Ontario, Canada in 1997 and M.S. and Ph.D. degrees in Electrical Engineering from Stanford University, Stanford, CA, in 1998 and 2002,  respectively. Since 2002, he has been with the Electrical and Computer Engineering Department at the University of Toronto, Toronto, Ontario, Canada, where he is now Professor and holds a Canada Research Chair (Tier 1) in Information Theory and Wireless Communications. His main research interests include information theory, optimization, wireless communications, and broadband access networks.

Prof. Wei Yu is a Fellow of the Canadian Academy of Engineering, and a member of the College of New Scholars, Artists and Scientists of the Royal Society of Canada. He received the Steacie Memorial Fellowship in 2015, the IEEE Marconi Prize Paper Award in Wireless Communications in 2019, the IEEE Communications Society Award for Advances in Communication in 2019, the IEEE Signal Processing Society Best Paper Award in 2017 and 2008, the Journal of Communications and Networks Best Paper Award in 2017, the IEEE Communications Society Best Tutorial Paper Award in 2015. He serves as the First Vice President of the IEEE Information Theory Society in 2020, and has served on its Board of Governors since 2015. He is currently an Area Editor for the IEEE Transactions on Wireless Communications, and in the past served as an Associate Editor for IEEE Transactions on Information Theory (2010-2013), as an Editor for IEEE Transactions on Communications (2009-2011), and as an Editor for IEEE Transactions on Wireless Communications (2004-2007). He served as the Chair of the Signal Processing for Communications and Networking Technical Committee of the IEEE Signal Processing Society in 2017-18. He was an IEEE Communications Society Distinguished Lecturer in 2015-16.
\end{IEEEbiography}

\end{document}